\documentclass[
superscriptaddress,
amsmath,amssymb,
aps, twocolumn, prb, floatfix
]{revtex4-2}

\usepackage[utf8]{inputenc}
\usepackage[T1]{fontenc}
\usepackage{mathptmx}
\usepackage{tabularx}
\usepackage[italian,english]{babel}
\usepackage[dvipsnames]{xcolor}
\usepackage[breaklinks,pdftex,pdfpagelabels,bookmarks]{hyperref}
\hypersetup{colorlinks=true,linktocpage=true,pdfstartpage=3,pdfstartview=FitH,breaklinks=true,
	pdfpagemode=UseNone,pageanchor=true,pdfpagemode=UseOutlines,plainpages=false,bookmarksnumbered, bookmarksopen=true,bookmarksopenlevel=1,hypertexnames=true,pdfhighlight=/0,urlcolor=Brown,linkcolor=MidnightBlue!60!Black,citecolor=PineGreen}
\usepackage[version=4]{mhchem}
\usepackage{physics}
\usepackage{siunitx}
\usepackage{comment}
\usepackage{graphicx}
\usepackage{dcolumn}
\usepackage{bm}
\usepackage{amsmath}
\usepackage{mathtools, nccmath}
\usepackage{amsfonts}
\usepackage{amssymb}
\usepackage{xcolor}
\usepackage{bbold}

\renewcommand{\vec}{\bm}

\begin{document}
	
	\preprint{APS/123-QED}
	
	\title{Andreev spin-noise detector}
	
	\author{R. Capecelatro}
	\email{roberto.capecelatro@unina.it} 
	\affiliation{Dipartimento di Fisica E. Pancini$,$ Università degli Studi di Napoli Federico II$,$ Monte S. Angelo$,$ via Cinthia$,$ I-80126 Napoli$,$ Italy}
	\author{V. Brosco}
	\affiliation{Institute for Complex Systems$,$ National Research Council and Dipartimento di Fisica$,$ Università ”La Sapienza”$,$ P.le A. Moro 2$,$ 00185 Rome$,$ Italy}
	\affiliation{Research Center Enrico Fermi$,$ Via Panisperna 89a$,$ 00184 Rome$,$ Italy}
	\author{G. Campagnano}
	\affiliation{CNR-SPIN$,$ UOS Napoli$,$ Monte S. Angelo$,$ via Cinthia$,$ I-80126 Napoli$,$ Italy}
	\author{P. Lucignano}
	\affiliation{Dipartimento di Fisica E. Pancini$,$ Università degli Studi di Napoli Federico II$,$ Monte S. Angelo$,$ via Cinthia$,$ I-80126 Napoli$,$ Italy}
	\begin{abstract}
	We investigate the possibility to employ magnetic Josephson junctions  as magnetic-noise detectors. To illustrate our idea, we consider a system consisting of a 
	quantum dot coupled to superconducting leads in the presence of an external magnetic field. Under appropriate assumptions,  we relate the noise in the Josephson current to magnetization noise. 
	At the magnetic field driven $0-\pi$ transition the junction sensitivity as magnetic noise detector is strongly enhanced and it diverges in the zero temperature limit. Moreover,  we demonstrate that, if also dot energy is affected by fluctuations, only the magnetic noise channel contributes to Josephson current noise response when the quantum dot is tuned in resonance with superconducting leads.
\end{abstract}

\maketitle

\section{Introduction}
Charge noise spectroscopy has long been recognized as a tool of crucial relevance to investigate the physics of quantum transport in mesoscopic systems and devices \cite{Landauer1998_Nature,Beenakker2003_PhysToday}.
 An example is provided by shot noise measurements used to demonstrate the fractional charge of quasiparticles in the fractional Quantum Hall effect \cite{Kane1994, de-Picciotto1997}. Recently, scanning tunneling microscopy approach for shot noise measurements revealed coherent electron tunneling from magnetic impurity into a s-wave superconductor via Yu-Shiba-Rusinov state \cite{Thupakula2022}.
 In this scenario, spin-noise spectroscopy (SNS) exploiting \emph{spin-fluctuations} to extract information about the system spin dynamics represents a promising investigation technique \cite{Sinitsyn2016}. 
Specifically, probing the electron spin dynamics we can gain information about the underlying microscopic interactions such as spin-orbit coupling and magnetic disorder \cite{Sinitsyn2016}.
 Among the several methods to probe magnetization fluctuations, Faraday rotation spectroscopy, whose working principle consists in measuring spin fluctuations from the Faraday polarization rotation of a linearly polarized light impinging on the sample, deserves to be mentioned \cite{Aleksandrov1981}.  
 In particular, this method has been successfully applied to study electron-spin dynamics for conduction electrons in bulk GaAs \cite{Oestreich2005}  and to extract spin relaxation time and Land\'{e} $g$ factor from the spin-noise signal. In Ref.\cite{Crooker2010}, electrons and holes $g$ factors have been derived for localized states in semiconductor (In, Ga)As/GaAs Quantum Dots (QDs) from the measured magnetic fluctuations. Furthermore, Faraday rotation spectroscopy of Quantum Dot Molecules (QDM) has allowed to resolve the coherent tunneling between QDs as well as the exchange-type spin-spin interactions \cite{Roy2013}, while it has been also used to study heterogeneous interacting spin systems via cross-correlation SNS \cite{Roy2015}.
Similarly, in view of having more compact experimental on-chip setups, SQUID (Superconducting Quantum Interference Device)-based magnetometry \cite{Barone1982, Tinkham} has been recognized as a valuable tool to measure magnetic field fluctuations in spin glasses \cite{Reim1986, Svedlindh1989} and superconductors \cite{Magnusson1998} with the sample being placed close to a dc-SQUID circuit \cite{Sinitsyn2016}.

    In this manuscript, we propose a spin-noise detector based on a Josephson device \cite{Josephson1962, DeGennes, ColemanBook2015, Zagoskin}.
	Recent experiments have demonstrated the possibility to fabricate ferromagnetic Josephson junctions having high quality factors and a plasma frequency in the GHz range \cite{Massarotti2015, Caruso2019, Asano2019, Minutillo2021, Ahmad2022, Ahmad2022PRB, Ahmad2023}. 
	These results hint at the possibility to employ these junctions to realize novel quantum devices and sensors, exploiting \textit{exchange} rather than \textit{orbital} magnetic phenomena \cite{Massarotti2015, Caruso2019, Asano2019, Minutillo2021, Ahmad2022, Ahmad2022PRB, Ahmad2023}.
	Our main idea is to exploit the magnetic field dependence of Andreev tunneling characterizing Josephson effect in ballistic quantum point contacts  \cite{ Glazman1989, Beenakker1991_A, Beenakker1991_B, Beenakker1992, Rozhkov1999, Vecino2003,Zazunov2003, Siano2004, Mahn-Soo2004,BergeretRodero2006, Benjamin2007,  Lee_Martin_PRL_2008, Meng2009_PRB, Rodero_Review_2011, Wentzell2016, Delagrange2016, Ke2019, Meden_2019,Whiticar2021} to detect spin fluctuations. We consider a single level Quantum Dot Josephson junction (SQDS JJ) in the presence of an external magnetic field and we show that, under appropriate conditions, the supercurrent noise fluctuations  can be directly related to spin noise.
    We identify the magnetic field induced $0-\pi$ transitions in SQDS JJs as the origin of enhanced current noise sensitivity to magnetic field fluctuations, that can be controlled by the system temperature, thus, suggesting a sizable amplification for magnetic noise even in the weak field limit \cite{Merkulov2002}. Moreover, also in the presence of dot energy fluctuations the SQDS JJ detector appears to be much more sensitive to spin rather than charge noise when the dot is tuned in resonance with the superconducting leads.
    
 The proposed device would provide the unique chance of accessing information about microscopic spin-noise sources from the knowledge of the junction equilibrium transport properties and it can be used to probe magnetization fluctuations in the ferromagnets employed as barriers in unconventional ferromagnetic Josephson junctions for superconducting qubits \cite{Yamashita2005, Yamashita2006, Yamashita2020, Ahmad2022,Ahmad2022PRB, Ahmad2023, Minutillo2021, Massarotti2015, Asano2019, Caruso2019}. Research interest in JJs with ferromagnetic barriers \cite{Ryazanov2001, Kontos2002, Sellier2004, Golubov2004, Buzdin2005, Robinson2006, Buzdin2003, Bergeret2005, Feofanov2010, Senapati2011, Bergeret2012}, as well as magnetic Quantum Dot junctions \cite{Rozhkov1999, Vecino2003, vanDam2006,DeFranceschi2010, BergeretRodero2006, Benjamin2007, Rodero_Review_2011,Sellier2005,Pal2018, Pal2019}, lies especially in their potential application as $\pi$-shifters \cite{Yamashita2005,Yamashita2006,Yamashita2020, DeFranceschi2010, vanDam2006, Delagrange2016, Ke2019, Siano2004}, in the possible implementation of tunable $0-\pi$ junctions \cite{Asano2019, Minutillo2021, Ahmad2022} and in the link between $0-\pi$ and topological parity transitions \cite{Marra2016, Marra2018, Whiticar2021}. In order to investigate the feasibility of employing ferromagnetic JJs in superconducting quantum circuits their magnetization noise characterization is needed. Furthermore, the investigation of noise in these systems  underlies the possibility to simultaneously highlight material specific features, such as the decay time of spin-spin correlations, and detect device properties.
    Interestingly enough, our work might be relevant also to model noise in a Josephson junction through a Yu–Shiba–Rusinov state, recently realized by Karan et al. \cite{Karan2022}, and the oscillations of Gilbert damping, recently revealed by Yao et al. in ferromagnetic Josephson junctions.\cite{yao2021}.
	
	The paper is structured as follows. In Sec.\ref{Sec_2_Model}, we present the SQDS JJ under study, together with the system and noise Hamiltonians, and we recall the Josephson current formula for equilibrium transport properties. In Sec.\ref{Sec_3_current_noise}, we describe the system current noise in the presence of static fluctuations in dot energy and magnetic field, and we identify the most favorable transport regime for magnetic noise detection from current fluctuations. In Sec.\ref{Sec_4_0_pi_transition_and_current_noise_amplification}, we recall the mechanisms underlying magnetic field driven $0-\pi$ transitions in SQDS JJs and recognize these $\pi$ switchings as a source of enhanced sensitivity of the current noise to magnetic fluctuations. Moreover, we point out the optimal working conditions to extract information about microscopic spin-noise sources from the equilibrium current noise. In Sec.\ref{Sec_5_current_noise_at_high_temperature}, we investigate how the Josephson current noise is modified by increasing the system temperature. Sec.\ref{Sec_6_Conclusions} summarizes our main findings.

\section{Model} \label{Sec_2_Model}

\begin{figure}[t]
	\centering
	\includegraphics[width=0.36\textheight,height=0.19\textheight]{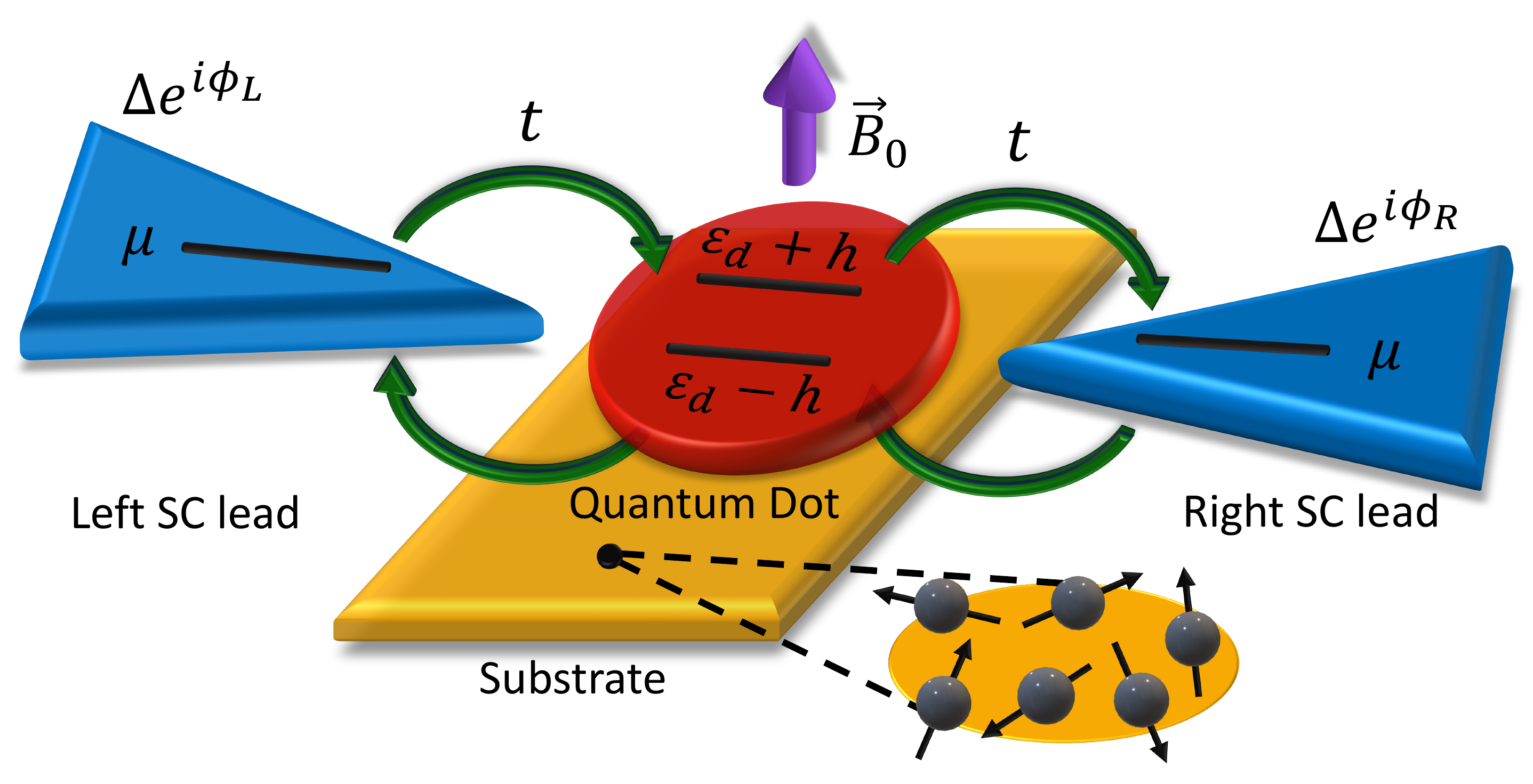}
	\caption{Scheme of the Superconductor - Quantum Dot -Superconductor Josephson junction (SQDS JJ) in the presence of an external magnetic field $\vec{B}_0$. Here, $\epsilon_{d}$ is the dot energy and $h=g\mu_B |\vec{B}_0|$ is the Zeeman splitting between the two spin channels affecting the dot level when $\vec{B}_0$ is turned on. The two s-wave superconductors are chosen to have equal gap $\Delta$ and chemical potential $\mu$. $\phi_{L/R}$ is the superconducting phase of the $L/R$ lead, respectively. $t$ is the amplitude of the hopping integral among the superconducting leads and the dot. Finally, a sketch of the fluctuations in the orientation of nuclear spins in the substrate beneath the dot is reported.}
	\label{Fig_1_QD_H_scheme}
\end{figure}

We model the device as a Josephson junction with a single level quantum dot (QD) barrier of energy $\epsilon_{d}$ in an external magnetic field $\vec{B}_0$, that we here chose to lie along the $z$-axis, thus perpendicular to the transport plane, i.e. $x-y$ plane, Fig.\ref{Fig_1_QD_H_scheme}.

In the presence of noise the Hamiltonian can be written as 
\begin{equation}
H=H_{\rm S}+H_{\rm noise}
\end{equation}
where $H_{\rm S}$  denotes the system Hamiltonian in the absence of fluctuations while  $H_{\rm noise}$ 
accounts for noise fluctuations.
The system Hamiltonian can be in turn cast as follows
\begin{equation}
	\label{Hamiltonian_system}
	H_{\rm S}=H_{\rm leads}+H_{\rm D}+H_{\rm T}\,,
\end{equation}
where $H_{\rm D}$, $H_{\rm leads}$ and $H_{\rm T}$ are the dot, the leads and the tunneling Hamiltonian, respectively.
The dot Hamiltonian, depending on the dot energy, $\epsilon_{d}$, and the Zeeman splitting provided by the magnetic field, $\vec{h}=h\vec{z}=g \mu_B  |\vec{B}_0|\vec{z}$, with $\mu_B$ and $g$ denoting the Bohr magneton and the electronic gyromagnetic ratio, reads
\begin{equation}
	H_{\rm D}=\epsilon_{d}\sum_{\sigma=\uparrow,\downarrow}d_{\sigma}^{\dagger}d_{\sigma}+h\left(d_{\uparrow}^{\dagger}d_{\uparrow}-d_{\downarrow}^{\dagger}d_{\downarrow}\right)\,,
	\label{H_Quantum_Dot}
\end{equation}
 where $d_{\sigma}$ indicates the annihilation operator for electrons of spin $\sigma=\uparrow,\downarrow$ on the dot.
 Following Ref.\cite{Benjamin2007}, we neglect Coulomb interaction on the dot. 
 For the sake of simplicity, the superconducting electrodes are supposed to be s-wave with equal chemical potential $\mu$, normal-state dispersion $\epsilon_{\vec{k},\sigma}$ and superconducting gap $\Delta$, thus yielding the following leads Hamiltonian
 \begin{equation}
 	\begin{split}
 		H_{\rm leads}=&	\sum_{i=L,R}\sum_{\vec{k}}\sum_{\sigma=\uparrow,\downarrow}(\epsilon_{\vec{k},\sigma}-\mu) c_{i,\vec{k},\sigma}^{\dagger}c_{i,\vec{k},\sigma}+ \\
 		&+\sum_{i=L,R}\sum_{\vec{k}} \Delta e^{i\phi_{i}}c_{i,\vec{k},\uparrow}^{\dagger}c_{i,\vec{k},\downarrow}^{\dagger} +\rm H.c. \hspace{0.3 mm},
 	\end{split}
 	\label{H_leads}
 \end{equation}
 where $c_{i,\vec{k},\sigma}$ represents the annihilation operator for electrons in the state $\vec{k}$ with spin $\sigma$ on the lead $i$ ($i=L,R$).
 Here, $\phi_{i}$ is the superconducting phase in the lead $i$, respectively. We set $\phi_L=-\phi/2$, $\phi_R=\phi/2$ and $\mu=0$.
 
 The tunneling Hamiltonian $H_{\rm T}$ reads
 \begin{equation}
 	H_{\rm T}=t\sum_{i=L,R}\sum_{\vec{k}}\sum_{\sigma}  c_{i,\vec{k},\sigma}^{\dagger}d_{\sigma} +\rm H.c. \hspace{0.3 mm},
 	\label{H_hopping}
 \end{equation}
 where the hopping amplitudes between the leads and the dot are chosen to be equal and $\vec{k}$-independent for both leads.
 
 When we assume rigid superconductors and constant hopping amplitude, noise Hamiltonian $H_{\rm noise}$ only involves dot energy and Zeeman field fluctuations, i.e. $\delta \epsilon_{d}$ and $\vec{\delta h}$, where the latter, in principle, can be non-collinear to the equilibrium exchange field $h\vec{z}$
\begin{equation}
		H_{\rm noise}= \delta\epsilon_{d}\sum_{\sigma=\uparrow,\downarrow}d_{\sigma}^{\dagger}d_{\sigma}+
		\sum_{\sigma, \sigma'=\uparrow,\downarrow}d_{\sigma}^{\dagger}\, (\hat{\vec \sigma} \cdot \vec{\delta h})d_{\sigma'}\,.
	\label{H_Dot_noise}
\end{equation}
Here, $\hat{\vec \sigma}=\left(\hat{\sigma_{1}},\hat{\sigma_{2}},\hat{\sigma_{3}}\right)$ is the vector of the Pauli matrices in the spin space.

Fluctuations in dot energy, $\delta \epsilon_{d}$, can originate from statistical retrapping processes of charge carriers in the substrate beneath the QD, possibly inducing fluctuations in the control gate voltage \cite{PellegrinoFalci2020, Pellegrino2021, Pellegrino2023, Paladino2014, Baladin2013}. Fluctuations of the Zeeman field may arise, in this geometry, due to the interactions between the dot electrons and spins of the nuclei in the substrate that, for weak external fields, can be described within the so-called "central spin model'' \cite{Merkulov2002, Sinitsyn2016,froehling2018}. In this framework, the intrinsic dynamics of the spin-bath happens on time-scales $\tau_i \simeq 100 \mu$s \cite{Sinitsyn2016} and thus can be neglected. The noise Hamiltonian only involves the hyperfine coupling between electrons and nuclei of the substrate, 
\begin{equation}
  H^{\rm hf}_{\rm noise}\simeq \sum_{n} \sum_{\sigma, \sigma'=\uparrow,\downarrow}d_{\sigma}^{\dagger}A_n \, (\hat{\vec \sigma} \cdot \vec  I_n)d_{\sigma'}
\end{equation}
where $\vec  I_n$ denotes the spin of the nucleus $n$ and $A_n$ quantifies the interaction between the n-th nucleus and the electron on the dot.

The effects of the hyperfine coupling between electrons and substrate nuclei, within the "frozen spin approximation'' \cite{Merkulov2002}, yields, for an ensemble of $\rm N$ nuclei, an effective Overhauser field given by $\vec B_{\rm N}= \sum_{i} A_i \, \vec  I_i$ \cite{Sinitsyn2016, Merkulov2002}. In this context, Zeeman field fluctuations would simply read as $\vec{\delta h}=\mu_B g \vec{\delta B_{\rm N}}$.
It can be shown that, within this approximation, due to the rotational symmetry of the system, the direction of the total Zeeman coupling, i.e. $h\hspace{0.2 mm} \vec{z}+\vec{\delta h}$, does not affect the junction transport properties. For this reason, later on we consider $\vec{\delta h}$ parallel to the equilibrium Zeeman field, i.e. $\vec{\delta h}=\delta h \hspace{0.1 mm} \vec{z}$.
\subsection{Josephson current}
 In the case of no bias voltage applied to the S electrodes, the equilibrium Josephson current, $J\left(\phi\right)$, is only driven by the phase difference $\phi$ between the leads and, in the Matsubara representation, can be written as follows \cite{Minutillo2021, Sellier2005, Benjamin2007, Meng2009_PRB,  Beenakker1992}
\begin{equation}
	\label{Current_Matsubara_Final}
		J\left(\phi\right)=  e T\Delta \Gamma \sin\left(\frac{\phi}{2}\right) \sum_{\omega_{n}}\frac{\Re\left(F_{dd,\downarrow\uparrow}(\omega_{n})\right)}{\sqrt{\Delta^2+\omega_{n}^2}} \; ,
\end{equation}
where $\omega_{n}=\pi\left(2n+1\right)T$ is the fermionic Matsubara frequency and $T$ is the system temperature, with $F_{dd,\downarrow\uparrow}(\omega_{n})$ being the $\downarrow\uparrow$ element of the anomalous dot Green's function (GF), describing the superconducting correlations on the dot. We approximate the two leads normal-state dispersion $\epsilon_{\vec{k},\sigma}$ by considering them as described by a flat and infinite band with a constant density of state $\rho_{0}$ (i.e. the leads density of states at the Fermi level) \cite{Benjamin2007, Sellier2005, Meng2009_PRB}, thus introducing the linewidth of the dot energy levels as simply $\Gamma=2\pi\rho_{0}t^2$, describing the dot-lead hybridization.
In this work, we set $\hbar=k_{B}=1$ and the Josephson current $J$ is scaled by $e\Delta$.

Since the line-width of the dot level $\Gamma$ plays the same role of Thouless energy $E_{Th}$ in diffusive SNS JJs \cite{Beenakker1991_A, Beenakker1991_B, Beenakker1992, Furusaki_Tsukada_1991, Furusaki_Takayanagi_Tsukada_1992}, the small and long junction conditions reading, respectively, $\Delta \ll E_{Th}$ and $\Delta \gg E_{Th}$, simply become $\Delta \ll \Gamma$ and $\Delta \gg \Gamma$ in SQDS JJs \cite{Beenakker1991_A, Beenakker1991_B, Beenakker1992}. Short junction limit is characterized by negligible quasiparticles contribution to Josephson current \cite{Beenakker1991_A, Beenakker1991_B, Beenakker1992}, thus representing the proper regime to simulate transport properties of novel tunnel ferromagnetic Josephson junctions with insulating barrier \cite{Minutillo2021, Ahmad2022, Massarotti2015,Ahmad2022PRB, Caruso2019, Asano2019}, more suitable for quantum circuits applications in view of the low quasiparticles current.
For this reason, in this work we analyze SQDS JJs characterized by $\Delta \ll \Gamma$ and all energies are scaled by $\Gamma$. 
In Appendix A, we calculate the dot GF when it is coupled to the leads, encoding information about the junction Andreev Bound States (ABS), through which the supercurrent flows \cite{Benjamin2007, Meng2009_PRB}, whose knowledge is necessary to compute the junction current-phase relation (CPR), i.e. $J\left(\phi\right)$.

\section{Current noise in SQDS JJs}
	\label{Sec_3_current_noise}
	We study the system current noise in the presence of both dot energy $\epsilon_{d}$ and Zeeman field $h$ fluctuations. Since we aim to exploit the junction current noise response as a probe of the magnetic noise source, we focus our attention on the intrinsic link between $J$ and $h$ fluctuations. In the following we assume the two noise sources to be uncorrelated.
	\begin{figure}[t]
		\centering
		\includegraphics[width=0.27\textheight,height=0.20\textheight]{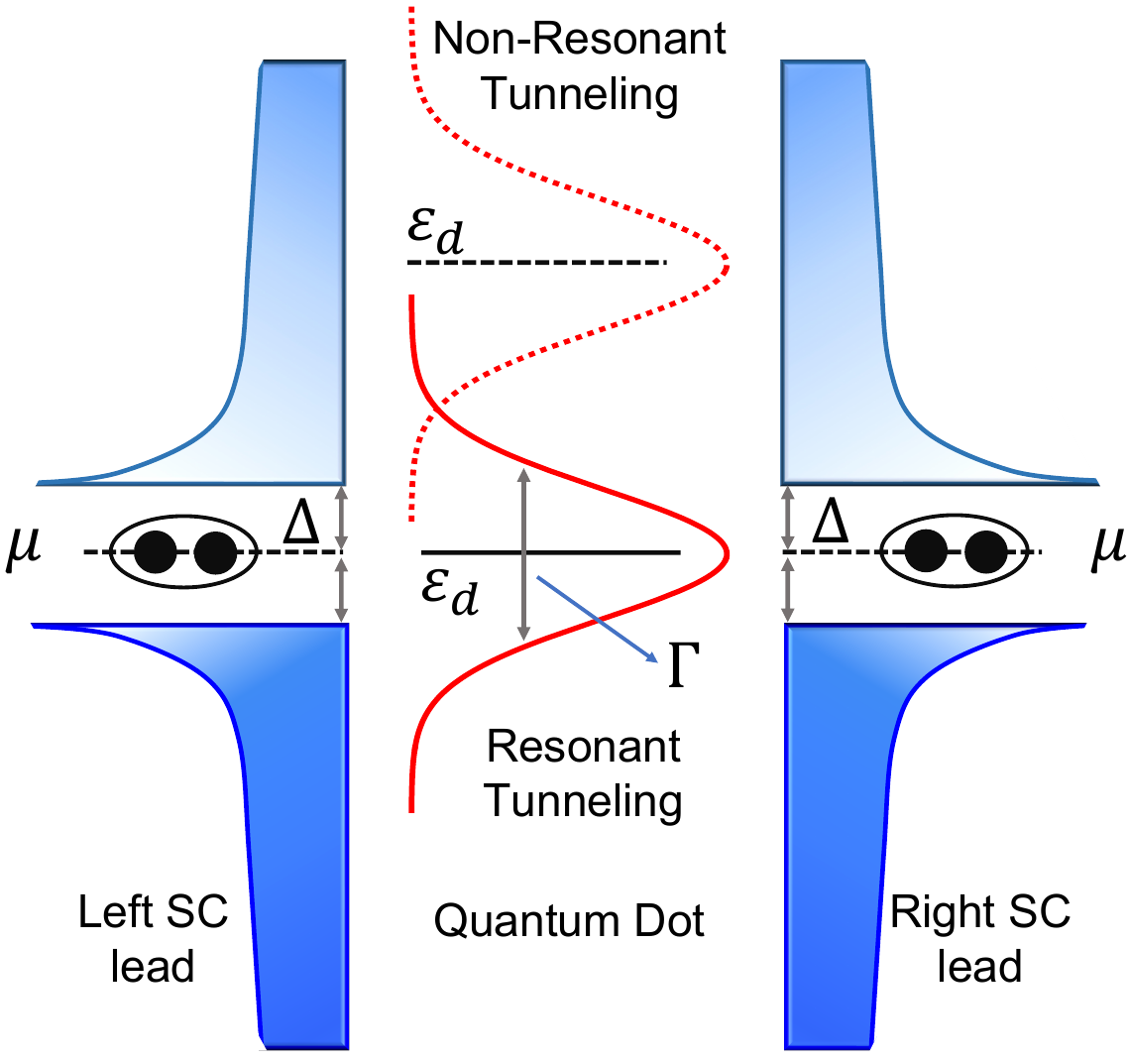}
		\caption{SQDS JJ level scheme in the absence of magnetic field. Dot level finite linewidth $\Gamma$ is the result of hybridization with the leads. By controlling $\epsilon_{d}$ one can reach both "resonant" and "non-resonant" tunneling regime.}
		\label{Fig_2_SQDS_levels_scheme}
	\end{figure}
	To keep the discussion simple, we further specify our analysis to the case of static Gaussian noise, where the QD degrees of freedom are characterized by time-independent zero-mean fluctuations, $\langle \delta h\rangle=\langle\delta\epsilon_{d}\rangle=0$, described by Gaussian probability distribution, $P\left(\delta \epsilon_{d}\right)$ and $P\left(\delta h\right)$, with variance given by $\sigma_{\epsilon_{d}}^2=\langle\left(\delta \epsilon_{d}\right)^2\rangle$ and $\sigma_{h}^2=\langle\left(\delta h\right)^2\rangle$, respectively.
	The assumption of Gaussian static noise can be justified for magnetic fluctuations considering that, in the central spin model and "frozen spin approximation'' \cite{Merkulov2002}, the Overhauser field $\vec B_{\rm N}$ yields classical static fluctuations, that in the limit of a large $\rm N$  are Gaussianly distributed with a standard deviation $\sigma_{B_{\rm N}} = B_{\rm N\,max}/\sqrt{N}$, $B_{\rm N\,max}$ indicating the maximum Overhauser field which is typically of the order of a few mT \cite{pal2017,chekhovich2013,froehling2018}.
	For an electron confined in a GaAs quantum dot and interacting with a typical number 100 spin-3/2 nuclei, this results in $\sigma_{B_{\rm N}} \sim 4$ mT \cite{chekhovich2013} and, consequently, in an overall magnetic field with probability distribution given by $P(B_0)=\exp(-B_0^2/(2\sigma_{B_{\rm N}}^2))/(\sqrt{2\pi}\sigma_{B_{\rm N}})$.
	A careful discussion of the limits of validity of the frozen spin approximation in external magnetic fields can be found in Ref.\cite{froehling2018}.	
  In the context of Gaussian static noise, the SQDS JJ current-noise characterization reduces calculate the Josephson current variance as a function of the superconducting phase difference between the leads can be expressed as \cite{Kittel1959, Kogan1996, Taylor1996}: 
 \begin{equation}
 \label{SigmaJ_Def_generic}
 \sigma_{J}^2\left(\phi\right)=\langle\left(\delta J(\phi)\right)^2\rangle=\langle J^2(\phi)\rangle-\langle J(\phi)\rangle^2
 \end{equation}   
 where $\langle \cdot \rangle$ is intended as the average over dot energy and Zeeman field fluctuations distributions, $P\left(\delta \epsilon_{d}\right)$ and $P\left(\delta h\right)$.
	
 		\begin{figure*} 
		\includegraphics[width=0.76\textheight, height=0.19\textheight]{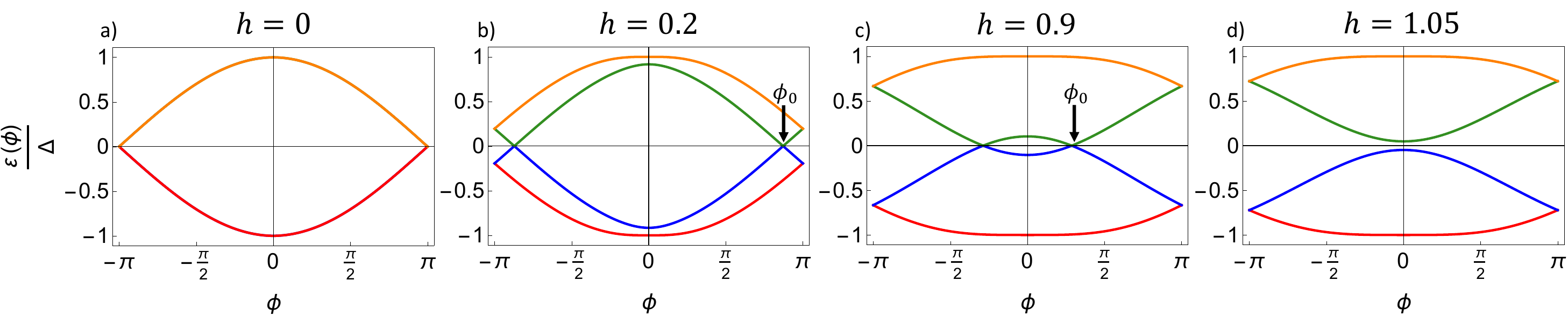}
		\caption[]{Andreev levels spectrum for the SQDS JJ ($\epsilon_{d}=0$, $\Gamma=1$, $\Delta=0.01$) along an Zeeman field induced $0-\pi$ transition, computed, from left to right, at $h=0$ (a), $h=0.2$ (b), $h=0.9$ (c) and $h=1.05$ (d), respectively. By increasing the Zeeman splitting between the different spin levels, Andreev Bound States (ABS) crossings at zero energy occur for $\phi=\pm\phi_{0}\neq0,\pm\pi$. Transition between the $0$ and $\pi$ phases is accomplished when the two near-zero ABS are switched. }
		\label{Fig_3_Andreev_lev_ed_0_all}
	\end{figure*}
 	In view of a possible application of the SQDS JJ as magnetic noise detector, we should be able to isolate and analyze the magnetic field fluctuations contribution to current noise response in Eq.\ref{SigmaJ_Def_generic}.
	In the presence of multiple noise sources this represents in general an hard task. 
	However, in the small fluctuations limit the current noise $\sigma_{J}$ can be approximated by the following expansion with respect to $\sigma_{\epsilon_{d}}$ and $\sigma_{h}$ \cite{Kittel1959, Kogan1996, Taylor1996, PellegrinoFalci2020}
	\begin{equation}
		\label{Sigma_J_sq_ed_h_noise}
		\begin{split}
			&\sigma_{J}^{2} \left(\phi\right)=\left(\frac{\partial J\left(\phi\right)}{\partial h}\bigg|_{\substack{\delta\epsilon_{d}=0\\ \delta h=0}}\right)^2 \sigma_ {h}^2 +\left(\frac{\partial J\left(\phi\right)}{\partial \epsilon_{d}}\bigg|_{\substack{\delta\epsilon_{d}=0\\ \delta h=0}}\right)^2 \sigma_ {\epsilon_{d}}^2 +\\
			&+\frac{1}{2} \left(\frac{\partial^2 J\left(\phi\right)}{ \partial\epsilon_{d}^2}\bigg|_{\substack{\delta\epsilon_{d}=0\\ \delta h=0}}\right)^2\sigma_{\epsilon_{d}}^4+\frac{1}{2} \left(\frac{\partial^2 J\left(\phi\right)}{ \partial h^2}\bigg|_{\substack{\delta\epsilon_{d}=0\\ \delta h=0}}\right)^2\sigma_{h}^4 - \\
			&-\frac{1}{2} \left(\frac{\partial^2 J\left(\phi\right)}{\partial \epsilon_{d}^2}\bigg|_{\substack{\delta\epsilon_{d}=0\\ \delta h=0}}\frac{\partial^2 J\left(\phi\right)}{\partial h^2}\bigg|_{\substack{\delta\epsilon_{d}=0\\ \delta h=0}}\right) \sigma_{\epsilon_{d}}^2\sigma_{h}^2,
		\end{split}
	\end{equation}
    where make the assumption of uncorrelated noise sources and express the fourth order moment of $\delta\epsilon_{d}$ and $\delta h$ as $\langle\left(\delta \epsilon_{d}\right)^4\rangle=3\sigma_{\epsilon_{d}}^4$ and $\langle\left(\delta h\right)^4\rangle=3\sigma_{h}^4$ \cite{Kittel1959, Kogan1996,Taylor1996}. As one can see in Eq.\ref{Sigma_J_sq_ed_h_noise}, the JJ current-noise response is determined by the current derivatives with respect to $\epsilon_{d}$ and $h$. The two contributions are in general difficult to separate.    
 The existence of "sweet spots" in the Hamiltonian parameters space where $\sigma_{\epsilon_{d}}$ contribution to $\sigma_{J}$ can be disregarded suggests a strategy to isolate the magnetic noise contribution. Specifically, when we tune the dot in resonance with the two superconducting electrodes (see Fig.\ref{Fig_2_SQDS_levels_scheme}) \cite{Beenakker1991_B, Beenakker1992}, i.e. $\epsilon_{d}=0$, $\partial_{\epsilon_{d}}J$ goes identically to $0$, actually predicting a vanishing first order contribution in $\sigma_{\epsilon_{d}}$ to $\sigma_{J}$. Thus, in case of negligible contribution from high order terms in $\sigma_{\epsilon_{d}}$ and $\sigma_{h}$, see Appendix B, the current variance $\sigma_{J}^2\left(\phi\right)$ in Eq.\ref{Sigma_J_sq_ed_h_noise} simply reduces to
	\begin{equation}
		\label{Sigma_J_sq_exp_h_noise_reduced}
			\sigma_{J}^{2} \left(\phi, \epsilon_{d}=0\right)=\left(\frac{\partial J\left(\phi, \epsilon_{d}=0\right)}{\partial h}\bigg|_{\substack{\delta\epsilon_{d}=0\\ \delta h=0}}\right)^2 \sigma_ {h}^2 ,
	\end{equation}
 To demonstrate that $\partial_{\epsilon_{d}}J$ vanishes at $\epsilon_{d}=0$ we recall the current formula in Eq.\ref{Current_Matsubara_Final} yielding $\partial_{\epsilon_{d}}J\propto\partial_{\epsilon_{d}}F_{dd,\downarrow\uparrow}$. 
 $F_{dd,\downarrow\uparrow}\left(\omega_{n}\right)$ has in general a cumbersome expression, that in the limit of zero Zeeman field $h=0$ \cite{Sellier2005, Meng2009_PRB} simplifies to
	\begin{equation}
		\begin{split}
			&F_{dd,\downarrow\uparrow}\left(\omega_{n}\right)=\frac{\Gamma  \Delta\cos\left(\frac{\phi}{2}\right)}{ 	\sqrt{\Delta ^2+\omega_{n}^2} } \times \\
			&\left(-\omega_{n}^2-\epsilon _{d}^2 -\frac{\Gamma ^2 \left(\Delta ^2 	\cos{\left(\phi/2\right)}^2+ \omega_{n}^2\right)}{\left(\Delta ^2+\omega_{n}^2\right)} -\frac{2 \Gamma \omega_{n}^2}{\sqrt{\Delta ^2+\omega_{n}^2}} \right)^{-1}
		\end{split} 
	\end{equation}
	Starting from the above equation the current derivative along $\epsilon_{d}$ can be easily calculated in the resonant tunneling where it identically vanishes
	\begin{equation}
    \partial_{\epsilon_{d}}F_{dd,\downarrow\uparrow}\bigg|_{\epsilon_{d}=0}=\partial_{\epsilon_{d}}J\bigg|_{\epsilon_{d}=0}=0 .
	\end{equation}
	This approach represents a promising starting point in view of detecting magnetic noise from the current variance and the knowledge of the junction equilibrium transport properties.
	Our purpose is to investigate in which conditions Eq.\ref{Sigma_J_sq_exp_h_noise_reduced} provides an accurate approximation to $\sigma_{J}^2\left(\phi\right)$, leading to a simple description of the system current noise response.
	In this scenario, essential for the practical applicability of this device as a magnetic noise detector is its sensitivity to $h$ fluctuations. For this reason, later on we concentrate on the current noise accompanying Zeeman field induced $0-\pi$ transitions, where the junction turns out to exhibit an increased $\sigma_{J}$ sensitivity to magnetic noise that is strictly connected to the choice of the quantum dot as the junction barrier.
	\section{$0-\pi$ transitions and current noise amplification in resonant SQDS JJs}
	\label{Sec_4_0_pi_transition_and_current_noise_amplification}
	\subsection{ Zeeman field driven $0-\pi$ transitions in SQDS JJs}
	The crucial role of the exchange field in controlling the switching between $0$ and $\pi$ phases in superconductor-ferromagnet-superconductor junctions is well established \cite{Ryazanov2001, Kontos2002, Sellier2004,Buzdin2005, Buzdin2003, Robinson2006,Golubov2004, Bergeret2005}. However, due to the peculiar nature of the barrier, it is useful to recall the mechanisms characterizing the $0-\pi$ transitions in Quantum Dot JJs \cite{Rozhkov1999,Vecino2003, BergeretRodero2006, Benjamin2007, Rodero_Review_2011}.
	\begin{figure}[b]
		\centering
		\includegraphics[width=0.33\textheight, height=0.20\textheight]{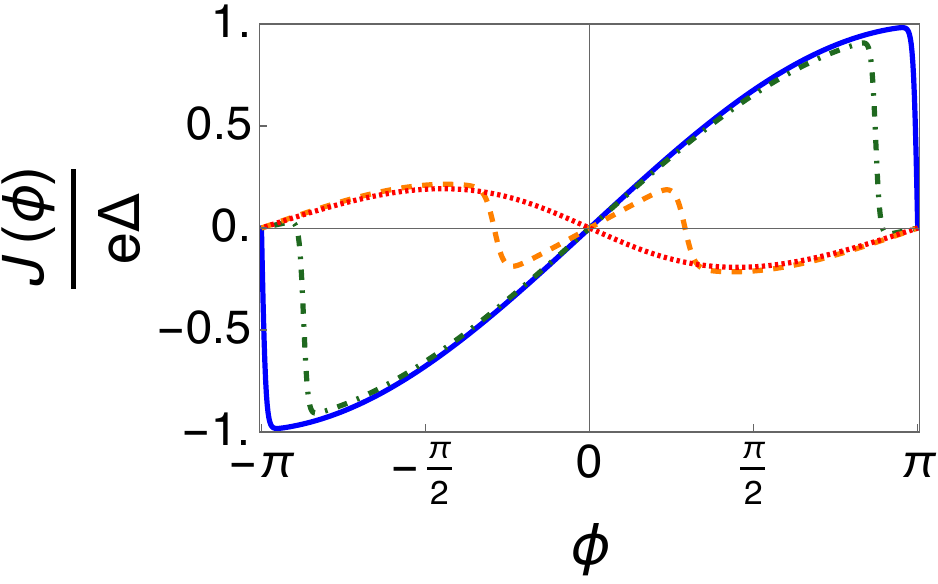}
		\caption{CPR behavior of the SQDS JJ in the resonant tunneling regime (at $\epsilon_{d}=0$, $\Delta=0.01$, $\Gamma=1$ and $T=0.02\hspace{1mm}T_{C}$) computed along the Zeeman induced $0-\pi$ switching at $h=0$ (blue solid curve), $h=0.2$ (green dot-dashed curve), $h=0.9$ (orange dashed curve) and $h=1.05$ (red dotted curve), respectively. Along the $0-\pi$ transition the JJ shows peculiar CPR jump discontinuities at the ABS crossings.}
		\label{Fig_4_CPR_comp_ed_0_h_var}
	\end{figure}
	For this purpose, in Fig.\ref{Fig_3_Andreev_lev_ed_0_all}, we show the ABS spectrum computed with increasing Zeeman interaction along an Zeeman field driven $0-\pi$, transition. It is useful to focus our attention on the absence of a gap in the ABS spectrum also when $h=0$, Fig.\ref{Fig_3_Andreev_lev_ed_0_all} (a), with Andreev levels crossings at $\phi=\pm\pi$ at zero-energy. Magnetic field affects the levels structure introducing a Zeeman splitting between different spin channels \cite{Rozhkov1999, Rodero_Review_2011}, Fig.\ref{Fig_3_Andreev_lev_ed_0_all} (b), with levels crossings appearing at $\phi=\pm\phi_{0}\neq\pm\pi$. By further increasing the dot Zeeman field the ABS crossing points are pushed toward $\phi_{0}=0$, Fig.\ref{Fig_3_Andreev_lev_ed_0_all} (c), thus, driving the system to the $\pi$ phase, where two particle-hole symmetric Andreev levels are switched, Fig.\ref{Fig_3_Andreev_lev_ed_0_all} (d). The crossings phase points $\pm\phi_{0}$ are, in principle, function of all the microscopic parameters entering the ABS levels calculation, i.e. $\epsilon_{d}$, $h$, $\Delta$ and $\Gamma$.

 	In Appendix C, we show that the presence of ABS crossings between particle-hole symmetric levels in this kind of SQDS JJ reflects in the occurrence of sharp jumps in the CPR , from positive to negative currents, at the crossings phases along $0-\pi$ transitions \cite{Vecino2003, Benjamin2007, Rozhkov1999, Rodero_Review_2011, Meden_2019, Marra2018}, as it is visible in Fig.\ref{Fig_4_CPR_comp_ed_0_h_var}, where the junction current-phase relations $J(\phi)$ corresponding to the ABS spectra in Fig.\ref{Fig_3_Andreev_lev_ed_0_all}, computed at temperature $T=0.02\hspace{1mm}T_{C}$, are reported. Peculiar jump-discontinuities in CPR, well known in ballistic quantum point contacts (QPCs) in resonant regime and short ballistic JJs \cite{Beenakker1991_B, Beenakker1992, Golubov2004, Rozhkov1999, Lee2015}, have been already predicted in SQDS junctions along the $0-\pi$ switchings \cite{Benjamin2007, Zazunov2003, Golubov2004} also in the presence of Coulomb interaction on the dot \cite{  Rozhkov1999, Vecino2003,Zazunov2003, Mahn-Soo2004,BergeretRodero2006, Benjamin2007,  Lee_Martin_PRL_2008, Meng2009_PRB, Wentzell2016, Meden_2019, Whiticar2021, Marra2018}, as well as in ferromagnetic Josephson devices with QPCs, e.g. SFcFS \cite{Golubov2004}, and correspond to a system state characterized by the contemporary presence of two minima in the total energy at $\phi=0$ and $\pi$ \cite{Benjamin2007, Golubov2004, Zazunov2003, Rozhkov1999} (see Appendix C).
	\begin{figure}[h]
		\centering
		\includegraphics[width=0.35\textheight, height=0.25\textheight]{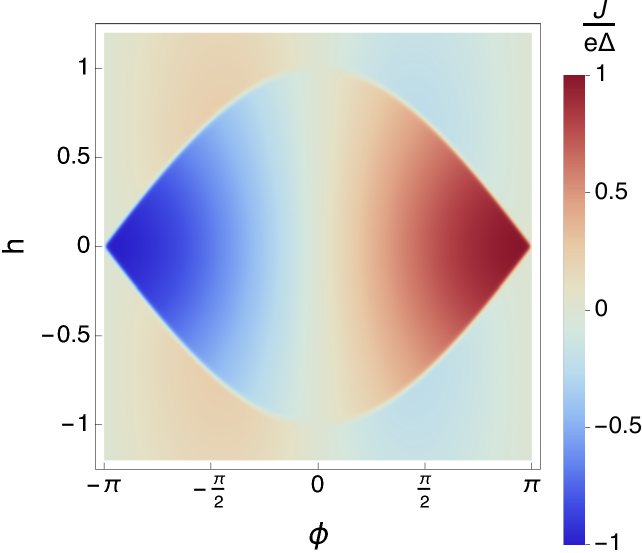}
		\caption{Density plot of Josephson current as a function of the phase and Zeeman field $J$($\phi$,$h$), computed at $\epsilon_{d}=0$, $\Gamma=1$, $\Delta=0.01$ and $T=0.02\hspace{1mm}T_{C}$. Transition between $0$ and $\pi$ phases is characterized by an intermediate transport regime in which the junction CPR shows peculiar jump-like discontinuities, e.g. abrupt change in the plot color when moving along the $\phi$ axis at fixed Zeeman field $h$ value.}
		\label{Fig_5_J_map_vs_phi_h_ed_0}
	\end{figure}
	This peculiar CPR behavior goes along with an enhanced contribution of higher harmonics, that is a well established signature of the transition to $\pi$ state in JJs with ferromagnetic barriers \cite{Buzdin2005, Bergeret2005, Asano2019, Ahmad2022, Minutillo2021}.
	We can better appreciate and summarize the transport properties of the SQDS junction by looking at Fig.\ref{Fig_5_J_map_vs_phi_h_ed_0}, where the density plot of the Josephson current as a function of the phase and Zeeman field, i.e. $J$($\phi, h$), (computed for $\epsilon_{d}=0$, $\Delta=0.01$ and $T=0.02\hspace{1mm}T_{C}$) is presented.
	Josephson current $J\left(\phi\right)$ is shown along the lines at fixed $h$ value on the $y$-axis. To accomplish the $0-\pi$ transition, the system must move across an intermediate regime in which the $J(\phi)$ shows sharp jump discontinuities, corresponding to sudden changes of color in the graphic. In the resonant tunneling regime, we can observe the lack of a region of the parameter space where the junction exhibits a pure $0$ behavior, starting its transitions toward the $\pi$ state as soon as the Zeeman coupling is turned on. $\pi$ switching is accomplished when $h=\pm1$ corresponding to the situations where both the Zeeman splitted dot levels with linewidth $\Gamma=1$ do not overlap anymore with the superconducting banks chemical potential $\mu_{s}=0$, thus having the junction exiting the resonant tunneling regime, (Figs.\ref{Fig_1_QD_H_scheme}-\ref{Fig_2_SQDS_levels_scheme}).

\subsection{Current noise amplification in the presence of Zeeman field fluctuations}

Occurrence of CPR jumps at the ABS crossing phases, $\phi=\pm\phi_{0}$, gains relevance when the Hamiltonian parameters on which $\phi_{0}$ depends are affected by noise. In our case, CPR discontinuities may lead to an increased sensitivity of the current noise along the $0-\pi$ transition to $\epsilon_{d}$ and $h$ fluctuations ($\Gamma$ and $\Delta$ are kept fixed), see Appendix C.

\begin{figure}[t]
	\centering
	\includegraphics[width=0.35\textheight, height=0.50\textheight]{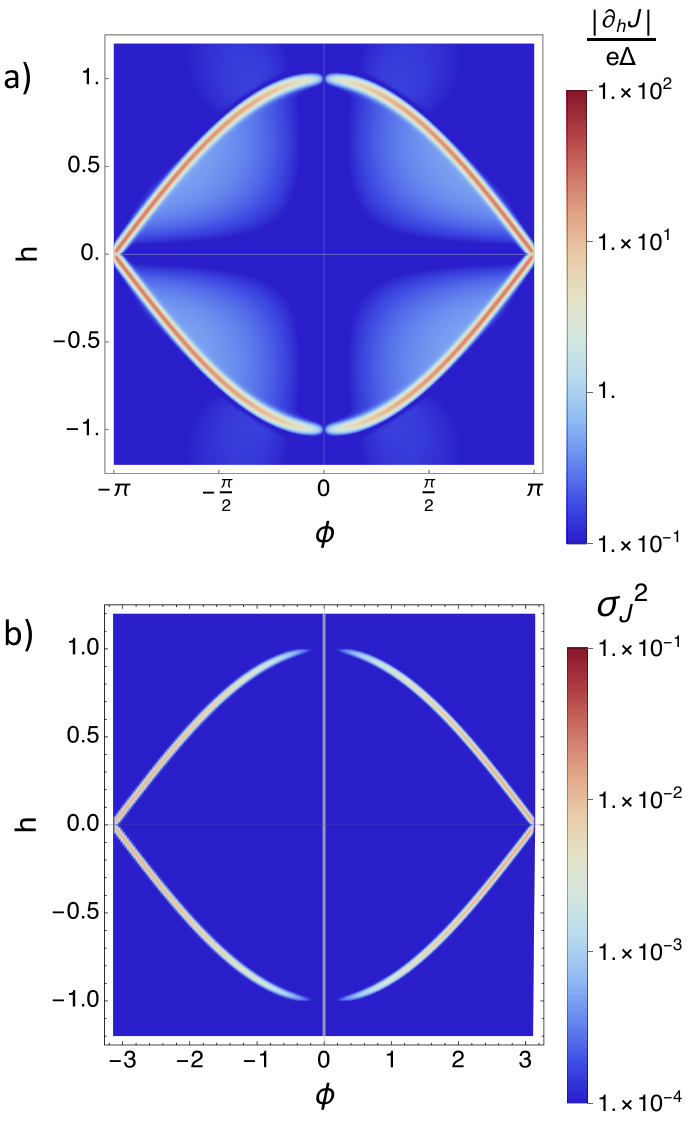}
	\caption{Density plot of the current derivative along $h$ (upper panel, (a)) and of the current variance $\sigma_{J}^{2}\left(\phi,h\right)$ (lower panel (b)) as a function of the superconducting phase difference and the Zeeman field ($\epsilon_{d}=0$, $\Delta=0.01$, $\Gamma=1$ and $T=0.02\hspace{1mm}T_{C}$). $\sigma_{J}^{2}\left(\phi,h\right)=\langle J^2\left(\phi\right)\rangle-\langle J\left(\phi\right)\rangle^2$ is computed in the presence of dot energy and Zeeman field fluctuations with equal standard deviation $\sigma_{\epsilon_{d}}=\sigma_{h}=0.005$. Along $0-\pi$ transition $\partial_{h}J$ divergences at the CPR jumps appear and the SQDS JJ exhibits current noise peaks.}
	\label{Fig_6_dJh_sigmaJsqh_map_vs_phi_h_ed_0}
\end{figure}

In particular, the abrupt change in sign of $J\left(\phi\right)$ leads to $\delta$-like peaks in the current derivatives with respect to the dot energy and Zeeman field, as it happens for $\partial_{h}J\left(\phi\right)$ shown in the density plot in Fig.\ref{Fig_6_dJh_sigmaJsqh_map_vs_phi_h_ed_0}(a). 
Here, the amplitude of $\partial_{h}J\left(\phi\right)$ along the $0-\pi$ switching is enhanced by two orders of magnitude at the ABS crossings with respect to the background.
\begin{figure}[h!] 
	\includegraphics[width=0.33\textheight, height=0.67\textheight]{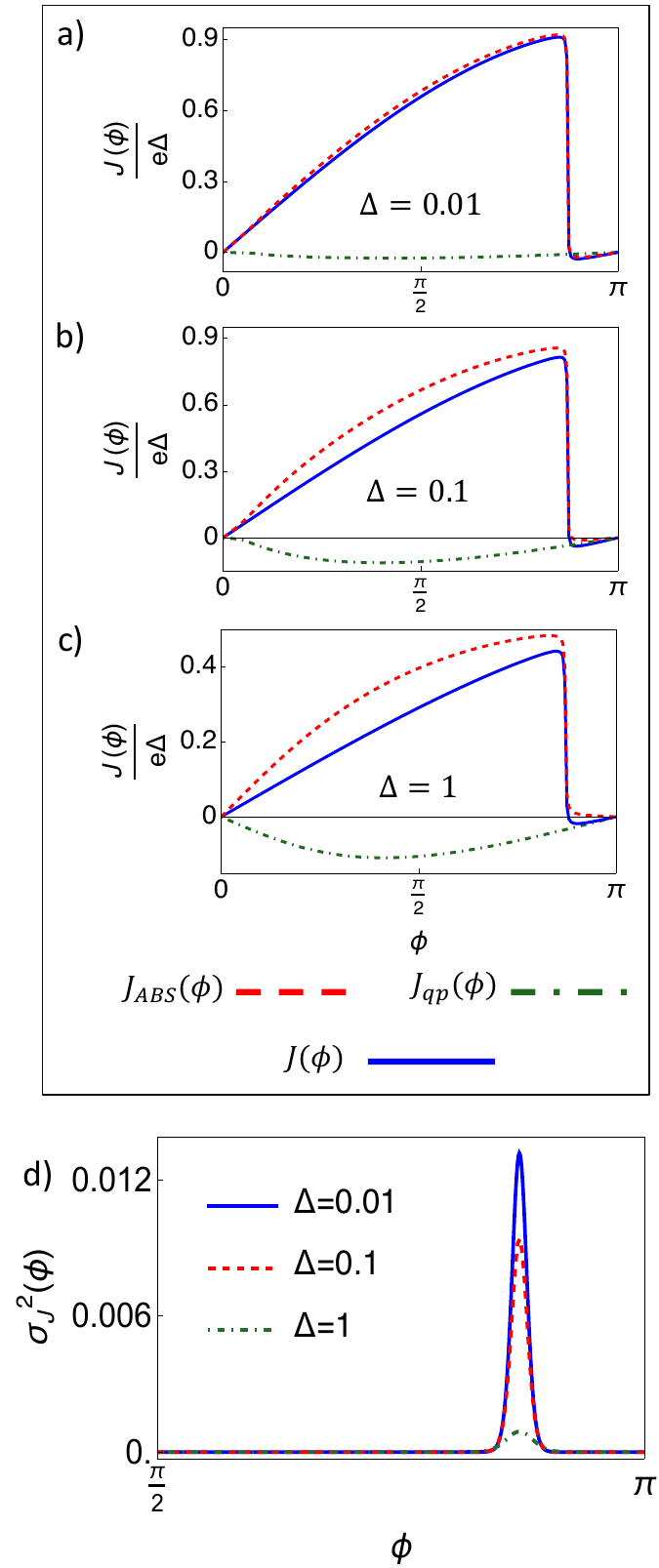}
	\caption[]{In the upper panel (a-c) the current-phase relation (blue solid lines) together with the relative current contributions of ABS (red dashed lines) and quasiparticles (green dot-dashed lines), in the lower panel (d) the current variance for $\sigma_{\epsilon_{d}}=\sigma_{h}=0.005$, computed at different values of the superconducting gap $\Delta$ (at $\epsilon_{d}=0$, $h=0.2$, $\Gamma=1$ and $T=0.02\hspace{1mm}T_{C}$), are respectively shown. As long as the system is in the short junction limit, i.e. $\Delta\ll\Gamma$, quasiparticles current remains negligible, while it reaches the same order of magnitude of the current carried by Andreev levels when $\Delta$ approaches $\Gamma$. Reduction of CPR jumps due to quasiparticles current leads to strong attenuation of the current noise peaks at the ABS crossings, when $\Delta$ approaches $\Gamma$, with the SQDS JJ exiting the short-junction regime }
	\label{Fig_7_CPR_sigmaJsq_ed_0_h_0_2_Deltavar}
\end{figure}
	\begin{figure*}
		\includegraphics[width=0.69\textheight, height=0.245\textheight]{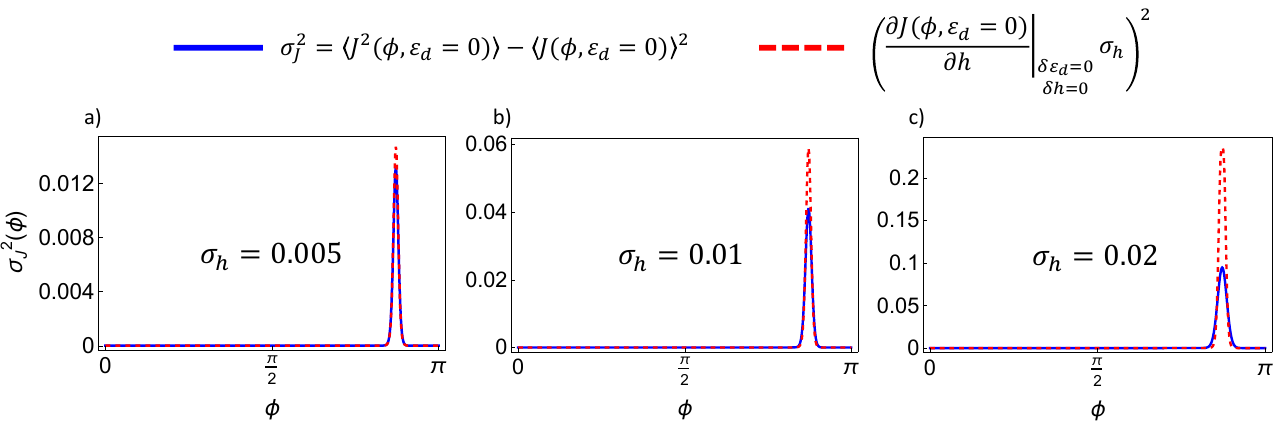}
		\caption[]{Comparison between $\sigma_{J}^{2}(\phi)$ calculated by the means of Eq.\ref{SigmaJ_Def_generic} and Eq.\ref{Sigma_J_sq_exp_h_noise_reduced} for the CPR curve at $h=0.2$ in Fig.\ref{Fig_4_CPR_comp_ed_0_h_var} ($\epsilon_{d}=0$, $\Gamma=1$, $\Delta=0.01$), at different values of the dot energy and Zeeman field standard deviations, $\sigma_{\epsilon_{d}}=\sigma_h=0.005$ (a), $\sigma_{\epsilon_{d}}=\sigma_h=0.01$ (b) and $\sigma_{\epsilon_{d}}=\sigma_h=0.02$ (c), respectively.}
		\label{Fig_8_sigmaJsq_ed_0_h_0_2_sigmahvar_freccia}
	\end{figure*}
	In the presence of Zeeman field fluctuations, as long as the system is in the small fluctuations regime and the expansion in Eq.\ref{Sigma_J_sq_exp_h_noise_reduced} is valid, $\partial_{h}J(\phi)$ divergences at ABS crossing phases suggests an enhanced current noise response to $h$ fluctuations. On the other hand, in the resonant dot case, divergences in $\partial_{\epsilon_{d}}J$ along Zeeman field induced $0-\pi$ transitions are prevented, since $\partial_{\epsilon_{d}}J$ is identically zero for $\epsilon_{d}=0$, possibly indicating vanishing first order $\sigma_{\epsilon_{d}}$ contribution to current fluctuations. 
	Confirmation of  the prediction of amplified current noise response can be found in Fig.\ref{Fig_6_dJh_sigmaJsqh_map_vs_phi_h_ed_0} (b) where we show the current variance $\sigma_{J}^2$ obtained from Eq.\ref{SigmaJ_Def_generic} in the presence of both $\epsilon_{d}$ and $h$ fluctuations taken with the same standard deviation $\sigma_{\epsilon_{d}}=\sigma_{h}=0.005$ for simplicity.
	In this density plot, we  observe the appearance of pronounced $\sigma_{J}^{2}$ peaks at the ABS crossing phases. 
	The large amplitude of the peaks, rising two orders of magnitude over background, indicates that Zeeman field driven $0-\pi$ transitions in QD JJs can be exploited to amplify and detect small magnetic fluctuations through current noise response. This enhances the practical potentialities of these systems for probing Zeeman field noise. 
	In addition, we observe in Fig.\ref{Fig_6_dJh_sigmaJsqh_map_vs_phi_h_ed_0} (b) that the current noise amplification accompanying the $0-\pi$ transition starts as soon as the magnetic field is switched on, and the largest current noise response is achieved at very low Zeeman coupling values. This suggests the possibility to have a high detection sensitivity without employing strong external magnetic fields that could in principle have detrimental effects on the superconducting leads.
	
	The inherent device sensitivity, whose fingerprint are the CPR jumps, is tightly connected to the Andreev spectrum dependence on the microscopic parameters, and it may be reduced in the presence of strong quasiparticles currents.
 	We thus analyze Andreev levels and quasiparticles current contributions separately in Fig.\ref{Fig_7_CPR_sigmaJsq_ed_0_h_0_2_Deltavar} by varying the superconducting gap, starting from a deeply short junction limit, $\Delta= 0.01\Gamma$ to the case where $\Delta=\Gamma$.
	If the short junction regime is characterized by a negligible quasiparticles contribution to $J\left(\phi\right)$, \ref{Fig_7_CPR_sigmaJsq_ed_0_h_0_2_Deltavar} (a-b), when we increase the $\Delta/\Gamma$ ratio quasiparticles current gradually grows until reaching the same order of magnitude of ABS supercurrent when $\Delta \approx\Gamma$, as it is evident in Fig.\ref{Fig_7_CPR_sigmaJsq_ed_0_h_0_2_Deltavar} (c). We can see that when quasiparticles contribution becomes sizable it is opposite in sign with respect to ABS current \cite{Benjamin2007, Krichevsky_PRB_2000}, hence, leading to a reduction of the CPR jump. Hence, also the maximum of $\partial_{h}J$ divergent peaks is decreased, since their height is proportional to the jump discontinuity in $J\left(\phi\right)$. This indicates a strong attenuation of the current noise sensitivity when the superconducting gap approaches the dot linewidth, as shown in Fig.\ref{Fig_7_CPR_sigmaJsq_ed_0_h_0_2_Deltavar} (d), further justifying the choice of working in the short junction limit in order to have also the current noise response maximized. In the followings, we fix $\Delta=0.01$, thus quenching the quasiparticles contribution to supercurrent that is detrimental for noise amplification.
 
    Once the most favorable transport regime for magnetic noise detection has been determined, the success in using the SQDS JJ magnetic noise detector relies on the accuracy of Eq.\ref{Sigma_J_sq_exp_h_noise_reduced}. The latter, in resonant tunneling case, not only allows us to access information about magnetic noise sources but it assures that dot energy noise does not contributes to $\sigma_{J}$.  It is thus important to identify the regime of validity of the small fluctuations approximation.
	For this purpose, in Fig.\ref{Fig_8_sigmaJsq_ed_0_h_0_2_sigmahvar_freccia} we show the current variance $\sigma_{J}^2$ computed at different widths of the dot energy and Zeeman field distributions, i.e. $\sigma_{\epsilon_{d}}$ and $\sigma_{h}$, by comparing the outcomes of Eq.\ref{SigmaJ_Def_generic} and of Eq.\ref{Sigma_J_sq_exp_h_noise_reduced}.
	    \begin{figure}[ht]
		\includegraphics[width=0.35\textheight, height=0.23\textheight]{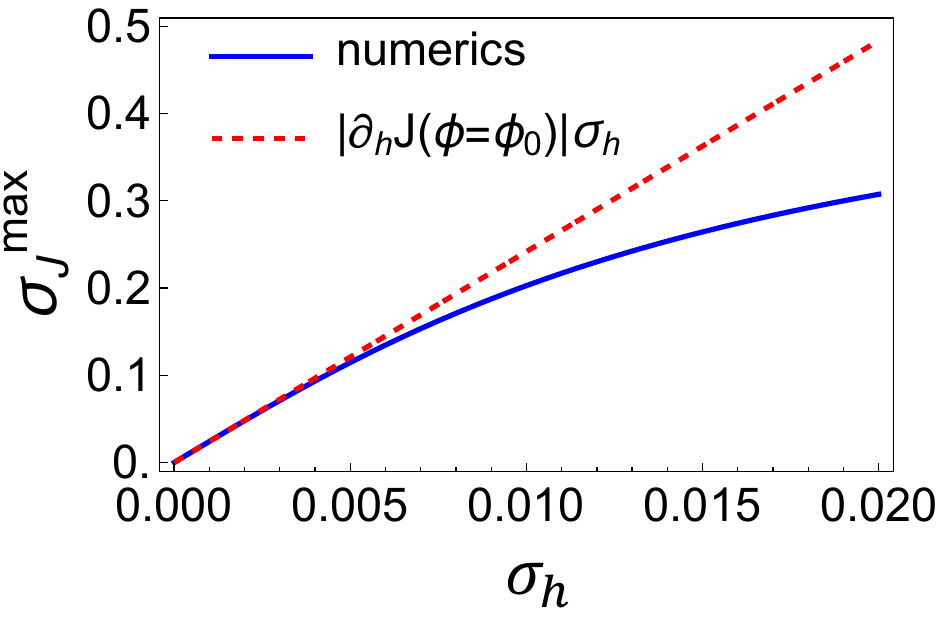}
		\caption{Maximum of the $\sigma_{J}$ peaks at the ABS crossings for the CPR curve at $h=0.2$ (green dot-dashed curve in Fig.\ref{Fig_4_CPR_comp_ed_0_h_var}) computed by varying the Zeeman field standard deviation $\sigma_{h}$ in comparison with the results predicted from first order $\sigma_{J}$ expansion in Eq.\ref{Sigma_J_sq_exp_h_noise_reduced}.}
		\label{Fig_9_SigmaJ_phi_max_vs_sigma_h}
	\end{figure}
	We note  that the smaller is $\sigma_{h}$, i.e. the narrower is the Zeeman field noise distribution, the higher is the precision of Eq.\ref{Sigma_J_sq_exp_h_noise_reduced} in reproducing the correct results both for $\sigma_{J}^2$ peaks height and width.
	These results are still better visualized when comparing the maximum of the $\sigma_{J}\left(\phi\right)$ peaks computed with the two approaches, in Fig.\ref{Fig_9_SigmaJ_phi_max_vs_sigma_h}, where we observe that for $\sigma_{h}\leq 5\times 10^{-3}$ Eq.\ref{Sigma_J_sq_exp_h_noise_reduced} succeeds in providing an accurate description for the current variance that hence turns out to be proportional to $\sigma_{h}$. Not only these results validate the small fluctuations expansion approach but they also demonstrates that dot energy noise can be disregarded for $\sigma_{J}^2$ calculation in the resonant tunneling.  In this scenario, these findings points out the possibility to extract information about the magnetic noise directly from the knowledge of the Josephson current variance and the junction equilibrium transport properties.
	Following Refs. \cite{Merkulov2002, Sinitsyn2016}, we highlight that the small fluctuations condition do not limit applicability of these systems as magnetic noise detectors. Indeed, we find that the expected Zeeman coupling distribution width $\sigma_{h}^{\rm exp}$ corresponding to magnetic field standard deviation $\sigma_{B_{\rm N}} \sim 4 \hspace{0.3 mm} \rm mT$, as mentioned in Refs.\cite{Merkulov2002, Sinitsyn2016}, is of the order of $\sigma_{h}^{\rm exp}\sim 10^{-7} \hspace{0.3 mm} \rm eV$, where we consider the electronic gyromagnetic factor in such heterostructures being $g\sim 0.3 \divisionsymbol 0.4$ \cite{Pryor2006,Pryor2006Erratum,Sheng2007,Stano2018, Camenzind2021}, while in our case, when dealing with systems such that $\Gamma \sim 100 \hspace{0.3 mm} \Delta$ and $\sigma_{h}=5\times10^{-3} \Gamma$, we predict Zeeman field standard deviation $\sigma_{h} = 5\times 10^{-5} \hspace{0.3 mm} \rm eV$ for $\Delta=10^{-1} \rm meV$.
	\section{High Temperature limit: temperature induced noise damping}\label{Sec_5_current_noise_at_high_temperature}
\begin{figure}[b]
		\centering
		\includegraphics[width=0.35\textheight, height=0.43\textheight]{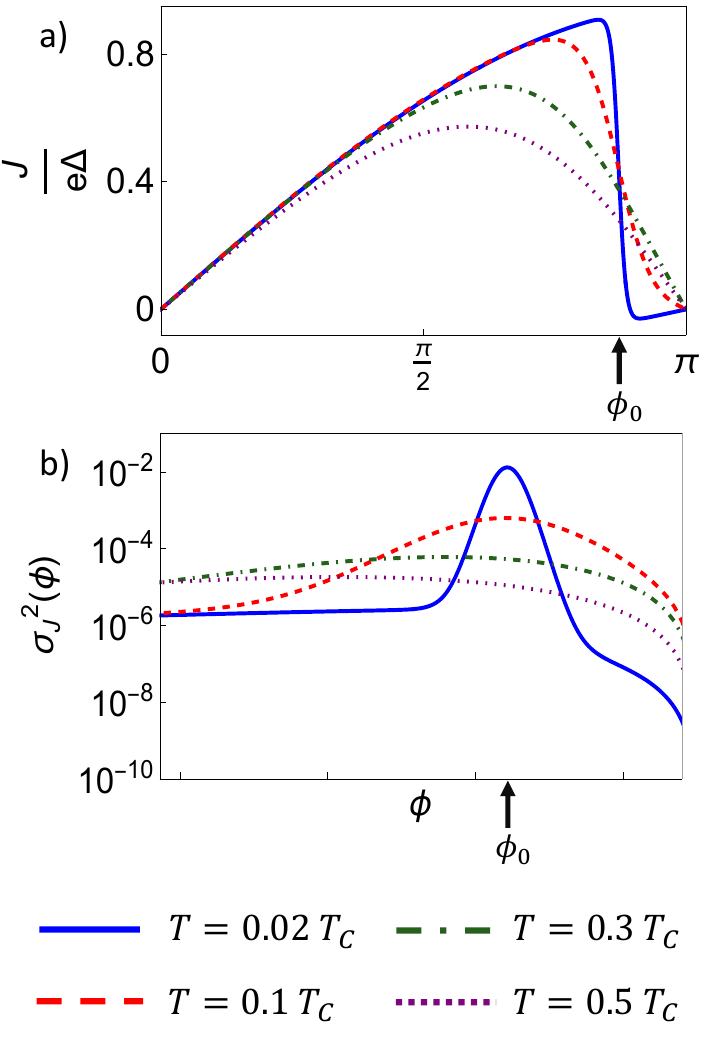}
		\caption{CPR and current variance $\sigma_{J}^2\left(\phi\right)$ modifications with the system temperature increasing. Results for the JJ with $h=0.2$ in Fig.\ref{Fig_4_CPR_comp_ed_0_h_var} computed at $T=0.02 \hspace{0.5mm}T_{C}$ (blue solid curve), $T=0.1 \hspace{0.5mm}T_{C}$ (red dashed curve), $T=0.3\hspace{0.5mm}T_{C}$ (green dot-dashed curve), $T=0.5\hspace{0.5mm}T_{C}$ (purple dotted curve), respectively, are reported. The temperature induced smoothing of CPR jumps and damping of the $\sigma_{J}$ noise peaks are remarkable.}
		\label{Fig_10_CPR_and_sigmaJsq_ed_0_h_0_2_T_var}
	\end{figure}

	In this section we study thermal effects on the sensitivity of the SQDS JJ detector. Starting from Eqs.\ref{Current_Matsubara_Final}-\ref{SigmaJ_Def_generic} we calculate the current and its noise by varying the system temperature in the range $\left[0.02 T_{C},0.5 T_{C}\right]$. The results for the QD JJ at $h=0.2$ (Fig.\ref{Fig_4_CPR_comp_ed_0_h_var}) are shown in Figs.\ref{Fig_10_CPR_and_sigmaJsq_ed_0_h_0_2_T_var} (a) and (b), respectively. We find that the sharp jumps, strongly evident when the temperature approaches zero, are significantly smoothed by the temperature.

    CPR jump discontinuities in such systems represent a fingerprint of the enhanced sensitivity to magnetic noise, these findings thus suggest a strong damping of the current noise peaks, as confirmed by Fig.\ref{Fig_10_CPR_and_sigmaJsq_ed_0_h_0_2_T_var} (b), where the amplitude of $\sigma_{J}$ peaks is reduced by one order of magnitude at $T=0.1\hspace{1mm}T_{C}$ and by more than two order of magnitude at $T=0.3\hspace{1mm}T_{C}$.

\begin{figure}[t]
		\includegraphics[width=0.34\textheight, height=0.46\textheight]{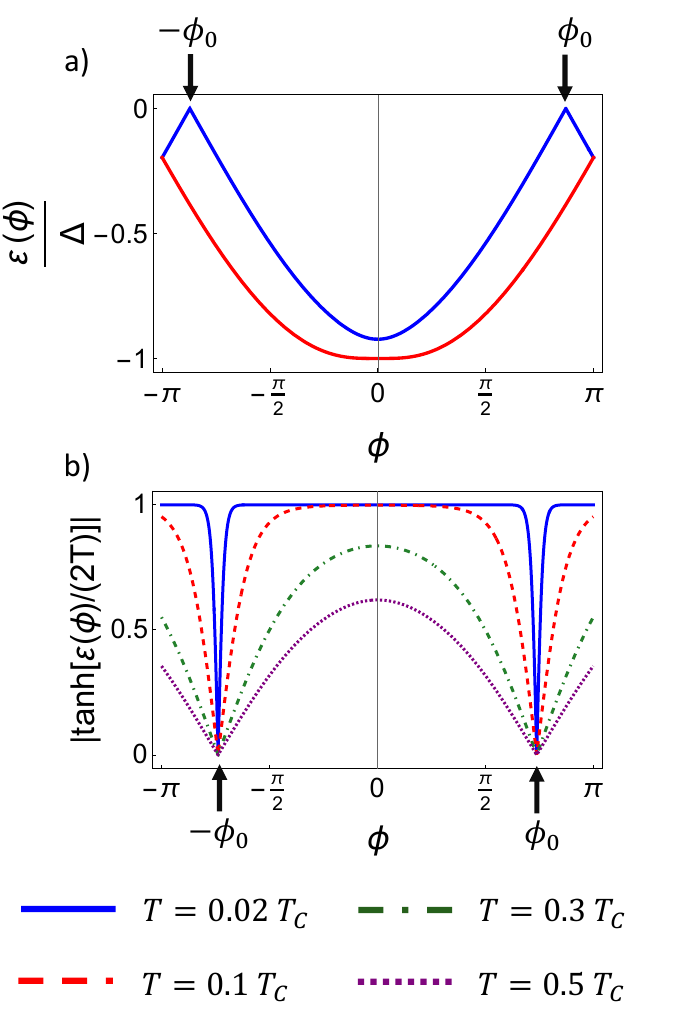}
		\caption{Negative energy states of the junction ABS spectrum (at $\epsilon_{d}=0$, $h=0.2$, $\Delta=0.01$ and $\Gamma=1$) lying below the superconducting leads chemical potential, through which supercurrent flows across the junction, (a). In (b) the thermal damping factor computed for the Andreev level exhibiting zero energy crossings at $T=0.02 \hspace{1mm} T_{C}$ (blue solid line), $T=0.1 \hspace{1mm} T_{C}$ (red dashed line), $T=0.3 \hspace{1mm} T_{C}$ (green dot-dashed line) and $T=0.5 \hspace{1mm} T_{C}$ (purple dotted line), respectively, is shown. A dip is present at the ABS crossing phases, whose width is enhanced by the temperature increasing, thus, providing the CPR jumps smoothing.}
		\label{Fig_11_Thermal_factor_ed_0_h_0_2_Delta_0_01_T_var}
	\end{figure}

	We can understand this effect by recalling the ABS current formula at finite temperature \cite{Beenakker1991_A,Beenakker1991_B, Beenakker1992, Furusaki_Tsukada_1991, Furusaki_Takayanagi_Tsukada_1992}
	\begin{equation}
		J_{ABS}\left(\phi\right)=\sum_{j, \epsilon\leq0}\frac{\partial \epsilon_{j}\left(\phi\right)}{\partial\phi}\tanh(\frac{\epsilon_{j}\left(\phi\right)}{2T}).
		\label{ABS_current_finite_T}
	\end{equation}
	and looking at the current carrying Andreev levels, i.e. at $\epsilon<0$, in Fig.\ref{Fig_11_Thermal_factor_ed_0_h_0_2_Delta_0_01_T_var}(a). When we compute the Boltzmann thermal factor for the Andreev state exhibiting zero energy crossings at $\phi=\pm\phi_{0}$, by varying the system temperature, Fig.\ref{Fig_11_Thermal_factor_ed_0_h_0_2_Delta_0_01_T_var}(b),
	we notice that it exhibits a visible dip at the crossings, whose width grows with the temperature. This leads to the lowering and widening of the CPR jumps, together with a strong reduction of $\sigma_{J}$ peaks. In this scenario, the SQDS JJ detector sensitivity is intrinsically related to the quantum mechanical nature of Andreev current governing the $0-\pi$ transition and thus is maximized at low working temperature, promising even improved performances when $T$ is further reduced

	\section{Conclusions and remarks}\label{Sec_6_Conclusions}
		In this work we studied the equilibrium current noise in Quantum Dot JJs in an external magnetic field in the presence of magnetic and dot energy fluctuations. We investigated the microscopic mechanisms underlying the current noise response to magnetic field fluctuations in these devices and to extract information about magnetic noise sources from current fluctuations, thus, paving the way to a novel spin-noise spectroscopy technique.
		We considered the case of static Gaussian noise, which is justified if the magnetic noise can be described in the "central spin model" within the "frozen spin approximation" \cite{Sinitsyn2016, Merkulov2002, froehling2018}.
		We identified the short junction regime and the resonant tunneling as the most favorable conditions for magnetic noise detection from the current noise response. Indeed,  we demonstrated that, when the dot is tuned in resonance with the leads, dot energy noise contribution to current fluctuations can be disregarded, thus leading to a simple proportionality relation between current variance $\sigma_{J}$ and the standard deviation of the dot Zeeman splitting $\sigma_{h}$, in the small fluctuations regime.
		Zeeman field induced $0-\pi$ transitions in Quantum Dot JJ turn out to amplify current noise response to magnetic fluctuations. In these systems, along $0-\pi$ switchings the occurrence of peculiar CPR jump discontinuities, intrinsically linked to the presence of Andreev levels crossings, is visible at low temperatures \cite{Benjamin2007, Vecino2003, Rodero_Review_2011}. CPR jumps are the origin of noise amplification, giving rise to strong current noise response, i.e. $\sigma_{J}\left(\phi\right)$, whose hallmarks are $\sigma_{J}$ peaks at ABS crossings. Although enhanced sensitivity of the current noise response to magnetic fluctuations along the $0-\pi$ transition may constitute a practical limit to employ these devices in quantum circuits, it also represents a unique opportunity to probe the magnetic fluctuations, accessing information about the microscopic noise sources from the junction equilibrium transport properties, thus, inspiring novel kind of Josephson spin-noise detectors. 
		In addition, we investigated the quasiparticles destructive influence on CPR jumps and on the detection sensitivity, pointing out the best working regime for maximizing the amplification of current response to magnetic noise. 
		In this scenario, system temperature $T$ plays a crucial role in smoothening the CPR jumps accompanying the $0-\pi$ transitions and washing out the $\sigma_{J}$ peaks, thus limiting the detectors sensitivity for $T \approx 10^{-1}\hspace{0.2 mm}T_{C}$ but suggesting even improved performances when $T$ is further reduced. For this reason, temperature turns out to be a valuable resource as a control knob of sensitivity to magnetic noise in these devices, also in view of their application in superconducting quantum circuits.
	\begin{acknowledgments}
		Financial support and computational resources from MUR, PON “Ricerca e Innovazione 2014-2020”, under Grant No. "PIR01\_00011 - (I.Bi.S.Co.)" are acknowledged. P.L. and V.B. acknowledge financial support from PNRR
		MUR project CN\_00000013-ICSC, PNRR MUR project PE0000023-NQSTI as well as from the project QuantERA II Programme STAQS project that has received funding from the European Union’s Horizon 2020 research and innovation programme under Grant Agreement No 101017733. The authors acknowledge E. Palladino, R. Fazio, A. Tagliacozzo  and M. Minutillo for fruitful discussions.
	\end{acknowledgments}
	
	\subsection*{Appendix A: Josephson current formula}\label{app_A}
	In this appendix, we present the calculation of the dot Green's function (GF) when it is connected to the superconducting leads by the means of perturbation theory. Moreover, we recall how to derive from its knowledge the Andreev levels spectrum and the Josephson current formula in Matsubara representation.
	
	Introducing the field operator formalism in Nambu$\otimes$spin space, we define $\psi_{D}$ and $\psi_{i,\vec{k}}$ as the field operators annihilating an electron on the QD and on the lead $i$ ($i=L,R$) in the state $\vec{k}$, respectively, as:
	\begin{equation}
		\psi_{D}=\begin{pmatrix}d_{\uparrow}, d_{\downarrow}, d^{\dagger}_{\uparrow}, d^{\dagger}_{\downarrow}\end{pmatrix}^T, \psi_{i,\vec{k}}=\begin{pmatrix}c_{i, \vec{k},\uparrow}, c_{i, \vec{k},\downarrow}, c^{\dagger}_{i,-\vec{k},\uparrow} c^{\dagger}_{i,-\vec{k},\downarrow}\end{pmatrix}^T.
		\label{Psi_Dot_leads}
	\end{equation}
	In this framework, leads, dot and tunneling Hamiltonians, respectively, are written as follows
	\begin{equation}
		\begin{split}
				\label{Hamiltonian_2nd_Q}
				H_{\rm leads}=&\sum_{i=L,R}\sum_{\vec{k}}\psi_{i,\vec{k}}^{\dagger}\check{H}_{i,\vec{k}}\psi_{i,\vec{k}}, \hspace{2 mm} H_{\rm D}=\psi_{D}^{\dagger}\check{H}_{\rm D}\psi_{D},\\
				&H_{\rm T}=\sum_{i=L,R}\sum_{\vec{k}}\psi_{D}^{\dagger}\check{H}_{\rm T}\psi_{i,\vec{k}}+\rm H.c. \hspace{0.2mm},
			\end{split}
	\end{equation}
	where $ \check{H}_{\rm D}$, $ \check{H}_{i,\vec{k}}$ and $ \check{H}_{T}$ are the QD, the lead $i$ and the hopping Hamiltonian matrices in Nambu$\otimes$spin space, respectively, reading
	\begin{equation}
		\begin{split}
		&\check{H}_{i,\vec{k}}=	(\epsilon_{\vec{k}_i}-\mu_{s})\hat{\sigma_{0}}\otimes\hat{\tau_{3}} + 	i\Delta e^{i\phi_{i}} \hat{\sigma}_{2} \otimes \left(\frac{\hat{\tau}_{1}+i\hat{\tau}_{2}}{2}\right) + \rm H.c. \hspace{0.2mm},\\
		&\check{H}_{\rm D}=\hat{\tau}_{3}\otimes(\epsilon_{d}\hat{\sigma}_{0}+h\hat{\sigma}_{3}), \hspace{1cm}\check{H}_{T}= t\hat{\sigma_{0}}\otimes\ \hat{\tau}_{3}
		 \hspace{0.5mm},
		\end{split}
	\end{equation}
	where we indicate with $\hat{\sigma}_{0}$ and $\hat{\sigma}_{\nu}$ ($\nu = 1, 2, 3$) the identity and the Pauli matrices in the spin space, respectively, while $\hat{\tau}_{0}$ and $\hat{\tau}_{\nu}$ ($\nu = 1, 2, 3$) play the same role in the Nambu space.
	 Here and in the following, the symbol $\hat{.}$ stands for $2 \times 2$ matrices in spin or Nambu space while $\check{.}$ indicates the $4 \times 4$ matrices in Nambu$\otimes$spin space.
	 
	From the expression of the system Hamiltonian in Nambu$\otimes$spin space, we can compute the dot Green's function when it is coupled to the superconducting leads, $\check{G}_{dd}$, whose structure is given by
	\begin{gather}
		\label{Nambu_spin_GF}
		\check{G}_{dd}=\begin{pmatrix}
			\hat{G}_{dd} & \hat{F}_{dd} \\
			-\hat{F}^{*}_{dd} & -\hat{G}^{*}_{dd}
		\end{pmatrix},
	\end{gather}
	with each block being a matrix in the spin space.
	The off-diagonal terms in the right-hand side of Eq.\ref{Nambu_spin_GF}  are the so-called \emph{anomalous} GFs $\check{F}_{dd}$, describing the superconducting pair correlations. Within the framework of perturbation theory, we can write a Dyson equation for $\check{G}_{dd}$ that, in the Matsubara representation, reads
	\begin{gather}
		\check{G}_{dd}(\omega_{n}) = \left(i\omega_{n}\check{1}-\check{H}_{\rm D}-\sum_{i=L,R} \check{H}_{\rm T}\check{G}^{0}_{i}(\omega_{n})\check{H}_{\rm T}\right)^{-1} \hspace{0.02 cm},
		\label{Gd_Dyson_equation_explicit}
	\end{gather}  
	where $\check{1}$ is the identity matrix in Nambu$\otimes$spin space, $\omega_{n}=\pi T(2n+1)$ is the fermionic Matsubara frequency with $T$ the system temperature.
	Here, $\check{G}^{0}_{i}(\omega_{n})$ is the \textit{bare} GF of the lead $i$ ($i=L,R$), that can be expressed in a simplified form, by assuming a constant density of state $\rho_{0}$ at the Fermi energy \cite{Benjamin2007, Sellier2005, Meng2009_PRB}:
	\begin{equation}
		\label{Bare_lead_GF}
		\begin{split}
			&\check{G}^{0}_{i}(\omega_{n})=\frac{-i\omega_{n}\pi \rho_{0}}{\sqrt{\Delta^2+\omega_{n}^2}}
			\check{1} +\\ 
			&+\frac{i\Delta\pi \rho_{0}}{\sqrt{\Delta^2+\omega_{n}^2}} \hat{\sigma}_{2} \otimes \left(e^{i\phi_{i}} \left(\frac{\hat{\tau}_{1}+i\hat{\tau}_{2}}{2}\right) - e^{-i\phi_{i}} \left(\frac{\hat{\tau}_{1}-i\hat{\tau}_{2}}{2}\right)\right).
		\end{split}
	\end{equation}
	
	By calculating the dot GF poles we access the knowledge about the junction ABS spectrum \cite{Zagoskin, ColemanBook2015}. Starting from the Eq.\ref{Gd_Dyson_equation_explicit}, and performing its analytic continuation, i.e. $i\omega_{n} \rightarrow z$, we obtain the Andreev Bound States (ABS) by solving in $z$ the following secular equation:
	\begin{equation}
		\det(z\check{1}-\check{H}_{D}-\sum_{i=L,R} \check{H}_{\rm T}\check{G}^{0}_{i}(z)\check{H}_{\rm T})=0.
		\label{ABS_det_eq}
	\end{equation}
	
	The Josephson current, driven by the phase difference $\phi=\phi_{R}-\phi_{L}$ established between the leads, can be calculated via the tunneling Hamiltonian method \cite{Barone1982,Zagoskin}. Without any loss of generality, we compute the Josephson current flowing from the dot to the right lead \cite{Zagoskin, Sellier2005}, that, in the thermal Matsubara representation \cite{Furusaki1994, Asano2001, Asano2019, Minutillo2021, Ahmad2022}, reads
	\begin{center}
		\begin{equation}
			\label{Current_Matsubara}
			J\left(\phi\right)= - \dfrac{i e}{2} T \sum_{\omega_{n}}Tr \left[ \hat{\sigma}_{0}\otimes\hat{\tau}_{3} \left(\check{H}_{\rm T}\check{G}_{c_{R},d} (\omega_{n}) -  \check{H}_{\rm T}\check{G}_{d,c_{R}} (\omega_{n})\right)   \right] \; ,
		\end{equation}
	\end{center}
	where $Tr$ stands for the trace over the Nambu$\otimes$spin space and  $\check{G}_{c_{R}d}$/$\check{G}_{dc_{R}}$ are the so-called lead-dot and dot-lead GFs describing the charge transfer between the dot and the right lead \cite{Zagoskin, Furusaki1994, Asano2001, Asano2019, Minutillo2021, Ahmad2022}. 
	Analogously to $\check{G}_{dd}$ in Eq.\ref{Gd_Dyson_equation_explicit}, also the GFs connecting leads and dot, can be calculated by the means of perturbation theory from the \emph{interacting} dot GF $\check{G}_{dd}$, Eq.\ref{Gd_Dyson_equation_explicit}, and \emph{bare}  right lead GF $\check{G}_{R}^{0}$. Their expression at the first non-vanishing order in the interaction ($\hat{H}_{\rm T}$) are, respectively: 
	\begin{equation}
		\begin{split}
			&\check{G}_{c_{R}d}\simeq \check{G}_{dd}\check{H}_{\rm T}\check{G}_{R}^{0},  \\
			&\check{G}_{dc_{R}}\simeq \check{G}_{R}^{0} \check{H}_{\rm T}\check{G}_{dd}.
		\end{split}
		\label{Lead-Dot_GFs}
	\end{equation}
	 By performing the trace over the Nambu space and exploiting the results in Eqs. \ref{H_hopping}, \ref{Nambu_spin_GF}, \ref{Current_Matsubara} and \ref{Lead-Dot_GFs}, the CPR formula can be further simplified
	\begin{center}
		\begin{equation}
			\label{Current_Matsubara_Simp}
			J\left(\phi\right)= \dfrac{i e t^{2}}{2} T \sum_{\omega_{n}}Tr_{\sigma} \left[\hat{F}_{dd} (\omega_{n}) \hat{F}_{R}^{0,*} (\omega_{n}) -  \hat{F}_{R}^{0} (\omega_{n}) \hat{F}_{dd}^{*} (\omega_{n})  \right] \; ,
		\end{equation}
	\end{center}
	where $ \hat{F}_{dd}$ and $\hat{F}_{R}^{0}$ are the anomalous blocks (in spin space) of the dot and bare R lead GFs, respectively, and the remaining trace $Tr_{\sigma}$  is over the spin space. It is worth noticing that only the superconducting correlation functions of the dot and bare R lead contribute to the charge transfer across the junction \cite{Sellier2005}. Eq.\ref{Current_Matsubara_Simp} can be rearranged in the more clear-cut formula in Eq.\ref{Current_Matsubara_Final} by performing the trace over spin space and exploiting both the expression for $\hat{F}_{s,R}^{0}$ in Eq.\ref{Bare_lead_GF} and the symmetries of dot anomalous GF in spin space, $\hat{F}\left(\omega_{n}\right)$.
	\appendix\subsection*{Appendix B: Small fluctuations expansion for current noise in resonant tunneling limit and different noise contributions}\label{app_B}
	In this appendix, we derive the formula for the current variance $\sigma_{J}^2\left(\phi\right)$ in Eq.\ref{Sigma_J_sq_ed_h_noise} for the resonant tunneling case. Moreover, we evaluate the different contributions to current noise expansion, Eq. \ref{Sigma_J_sq_ed_h_noise}, in order to demonstrate that, as long as the system is in the small fluctuations limit, the only non negligible contribution is provided by the first order term in $\sigma_{h}$, thus validating Eq.\ref{Sigma_J_sq_exp_h_noise_reduced}.
	
	When dot energy is tuned in resonance with leads chemical potential $\mu=0$, $\partial_{\epsilon_{d}}J|_{\epsilon_{d}=0}$ identically vanishes also when $h\neq0$.
	In this configuration, by assuming that the two noise channels, $\delta\epsilon_{d}$ and $\delta h$, are totally uncorrelated, for small fluctuations $J\left(\phi\right)$ can be expanded up to the second order as follows
	\begin{equation}
		\begin{split}
			\label{J_exp_resonant_Dot} 		
			&J\left(\phi,\epsilon_{d}=0\right)\approx J\left(\phi,\epsilon_{d}=0\right)\bigg|_{\substack{\delta\epsilon_{d}=0\\\delta h=0}}+\frac{\partial J\left(\phi,\epsilon_{d}=0\right)}{\partial h}\bigg|_{\substack{\delta\epsilon_{d}=0\\\delta h=0}}\delta h\\
			&+\frac{1}{2}\frac{\partial^2 J\left(\phi,\epsilon_{d}=0\right)}{\partial h^2}\bigg|_{\substack{\delta\epsilon_{d}=0\\\delta h=0}} \delta h^2+\frac{1}{2}\frac{\partial^2 J\left(\phi,\epsilon_{d}=0\right)}{\partial \epsilon_{d}^2}\bigg|_{\substack{\delta\epsilon_{d}=0\\\delta h=0}} \delta\epsilon_{d}^2 ,			
		\end{split}
	\end{equation}
	 yielding the following expression for the current fluctuations $\delta J=J-\langle J\rangle$
	\begin{equation}
		\begin{split}
			\label{delta_J_resonant_Dot} 		
			&\delta J\left(\phi,\epsilon_{d}=0\right)\approx \frac{\partial J\left(\phi,\epsilon_{d}=0\right)}{\partial h}\bigg|_{\substack{\delta\epsilon_{d}=0\\\delta h=0}}\delta h+\\
			&+\frac{1}{2}\frac{\partial^2 J\left(\phi,\epsilon_{d}=0\right)}{\partial h^2}\bigg|_{\substack{\delta\epsilon_{d}=0\\\delta h=0}} \left(\delta h^2-\langle \delta h^2\rangle \right)+\\
			&+\frac{1}{2}\frac{\partial^2 J\left(\phi,\epsilon_{d}=0\right)}{\partial \epsilon_{d}^2}\bigg|_{\substack{\delta\epsilon_{d}=0\\\delta h=0}} \left(\delta\epsilon_{d}^2-\langle \delta\epsilon_{d}^2\rangle \right),			
		\end{split}
	\end{equation}
	where we use that $\langle \delta \epsilon_{d}\rangle=0$ and $\langle \delta h\rangle=0$. Therefore, small fluctuations expansion for $\sigma_{J}^2\left(\phi\right)=\langle\delta J^2\left(\phi\right)\rangle$ reads
	
	\begin{equation}
		\label{Sigma_J_sq_ed_0_ed_h_noise}
		\begin{split}
			&\sigma_{J}^{2} \left(\phi,\epsilon_{d}=0\right)=\langle\delta J^2\left(\phi,\epsilon_{d}=0\right)\rangle=\\
                &\left(\frac{\partial J\left(\epsilon_{d}=0\right)}{\partial h}\bigg|_{\substack{\delta\epsilon_{d}=0\\\delta h=0}}\right)^2 \sigma_ {h}^2 +
			\frac{1}{2} \left(\frac{\partial^2 J\left(\epsilon_{d}=0\right)}{      \partial\epsilon_{d}^2}\bigg|_{\substack{\delta\epsilon_{d}=0\\\delta h=0}}\right)^2\sigma_{\epsilon_{d}}^4\\
                &+\frac{1}{2} \left(\frac{\partial^2 J\left(\epsilon_{d}=0\right)}{\partial h^2}\bigg|_{\substack{\delta\epsilon_{d}=0\\\delta h=0}}\right)^2\sigma_{h}^4 - \\
			&-\frac{1}{2} \left(\frac{\partial^2 J\left(\epsilon_{d}=0\right)}{\partial \epsilon_{d}^2}\bigg|_{\substack{\delta\epsilon_{d}=0\\\delta h=0}}\frac{\partial^2 J\left(\epsilon_{d}=0\right)}{\partial h^2}\bigg|_{\substack{\delta\epsilon_{d}=0\\\delta h=0}}\right) \sigma_{\epsilon_{d}}^2\sigma_{h}^2,
		\end{split}
	\end{equation}
	relating the CPR variance to the width of $\epsilon_{d}$ and $h$ statistical distributions, i.e. 
	$\sigma_{\epsilon_{d}}$ and $\sigma_{h}$.
	
	\begin{figure}[h]
		\includegraphics[width=0.36\textheight, height=0.38\textheight]{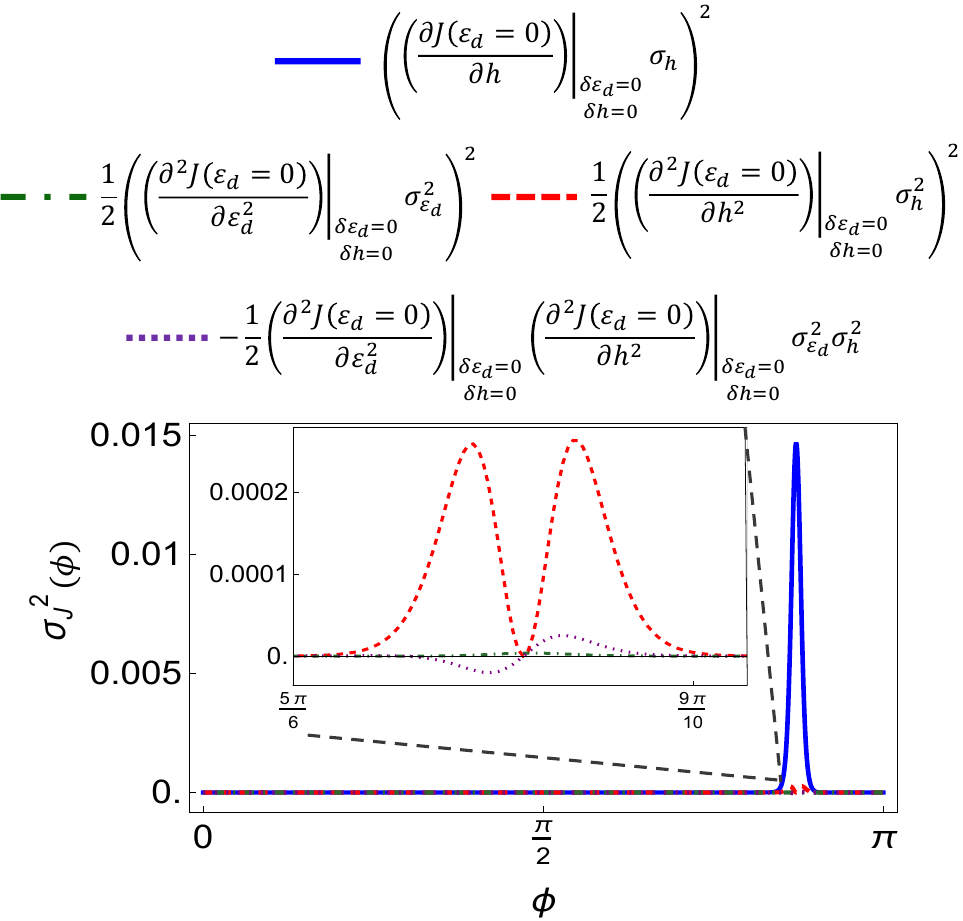}
		\caption{Different contributions to the small fluctuations expansion formula for the current variance $\sigma_{J}^2$, in Eq. \ref{Sigma_J_sq_ed_0_ed_h_noise} computed for the CPR curve at $h=0.2$ in Fig.\ref{Fig_4_CPR_comp_ed_0_h_var} ($\epsilon_{d}=0$, $\Gamma=1$, $\Delta=0.01$, $T=0.02 T_{C}$), for $\sigma_{\epsilon_{d}}=\sigma_{h}=0.005$, testifying that when the Quantum Dot is in resonance with the S leads the dot energy noise contribution to current noise is negligible.}
		\label{Fig_12_SigmaJ_contrib_exp_ed_0_h_0_2}
	\end{figure}
	
	In order to demonstrate that Eq.\ref{Sigma_J_sq_ed_0_ed_h_noise} reduces to Eq.\ref{Sigma_J_sq_exp_h_noise_reduced} it is necessary to evaluate the different contributions to CPR noise expansion, in order to investigate whether the dot energy fluctuations can be disregarded with respect to magnetic noise. For this purpose, in Fig.\ref{Fig_12_SigmaJ_contrib_exp_ed_0_h_0_2}, we report the different $\sigma_{J}$ series terms corresponding to the current variance in Fig.\ref{Fig_8_sigmaJsq_ed_0_h_0_2_sigmahvar_freccia} for $\sigma_{\epsilon_{d}}=\sigma_{h}=0.005$, where the small fluctuations expansion succeeds in providing a good description of current noise.
	We observe that the second order term in $\sigma_{h}$ is two order of magnitude lower than the first order one, while the latter exceeds terms involving $\sigma_{\epsilon_{d}}$ by at least three order of magnitude, thus, testifying that dot energy contribution to current noise can be considered negligible in the resonant tunneling limit and further proving the accuracy of Eq.\ref{Sigma_J_sq_exp_h_noise_reduced} in the small fluctuations regime.
	Nevertheless, it is worth noticing that the larger is $\epsilon_{d}$ standard deviation with respect to $\sigma_{h}$ the more significant higher order terms in $\sigma_{\epsilon_{d}}$ are for CPR noise.
 
	\appendix\subsection*{Appendix C: From Andreev levels crossings to jumps in CPR and divergences in current derivatives}\label{app_C}
\begin{figure}[t]
		\includegraphics[width=0.34\textheight,height=0.25\textheight]{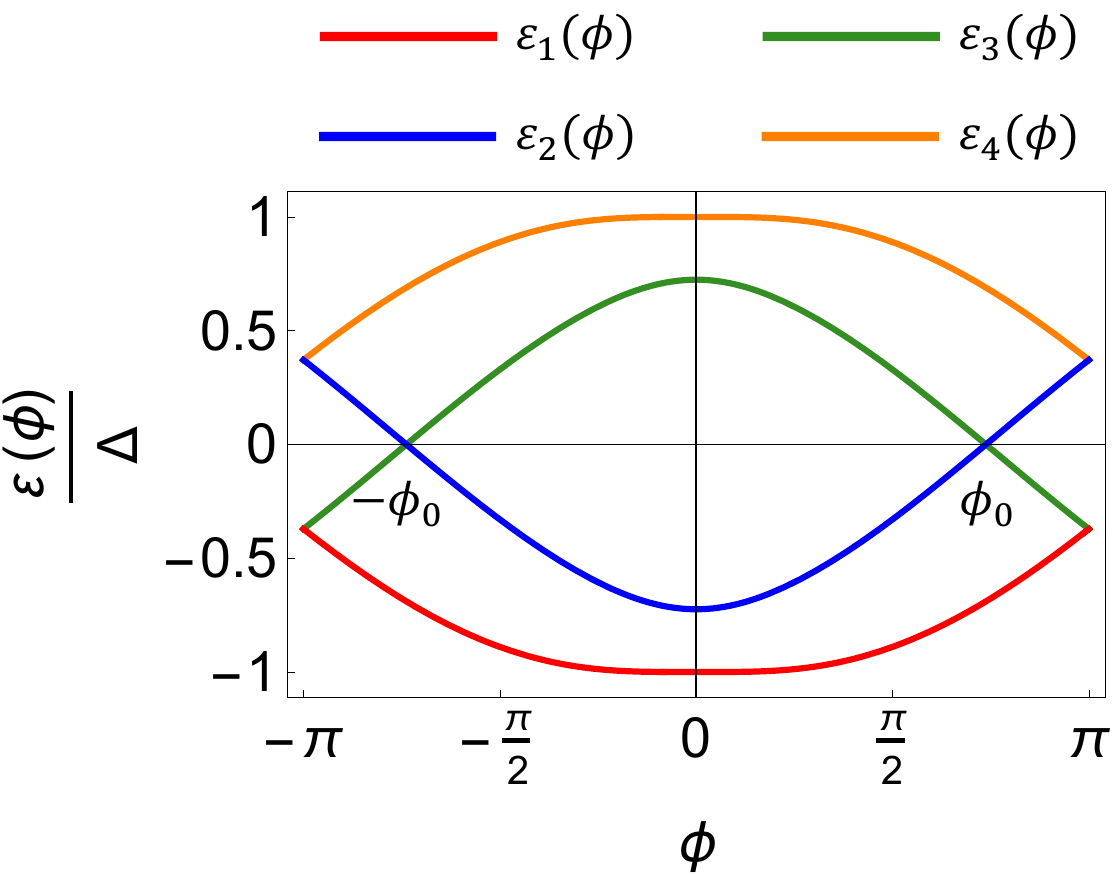}
		\caption{Example of Andreev levels crossings at $\pm \phi_{0} \neq \pm \pi$, for the SQDS JJ at $\epsilon_{d}=0$, $h=0.4$, $\Gamma=1$ and $\Delta=0.01$.}
		\label{Fig_13_A_1_Andreev_lev_scheme_ed_0_h_0_4}
	\end{figure}

	In this appendix, we provide a simple explanation to the connection between the crossings of level pairs in Andreev Bound States (ABS) spectrum and the jumps discontinuities in the Josephson current. 
	For the sake of clarity, in Fig.\ref{Fig_13_A_1_Andreev_lev_scheme_ed_0_h_0_4} we report as an example the ABS spectrum for the SQDS JJ in the resonant tunneling regime (at $\epsilon_{d}=0$, $h=0.4$, $\Gamma=1$ and $\Delta=0.01$) showing ABS crossings at $\pm \phi_{0} \neq \pm \pi$ and energy $\epsilon=0$.
	
	If the Andreev levels $\{\epsilon_{n}\}$ (with $n=1,...,4$) together with the crossing phases $\pm \phi_{0}$ are functions of the system microscopical parameters $\{\lambda_{i}\}=\{\epsilon_{d}, h, \Gamma, \Delta\}$, i.e. $\epsilon_{n}=\epsilon_{n}(\{\lambda_{i}\})$ and  $\phi_{0}=\phi_{0}(\{\lambda_{i}\})$, we clarify why the first current derivatives along $\{\lambda_{i}\}$, i.e. $\partial_{\lambda_{i}}J\left(\phi\right)$, show $\delta$-like divergences in correspondence of the levels crossings. 

\begin{figure}[b]
		\includegraphics[width=0.31\textheight,height=0.27\textheight]{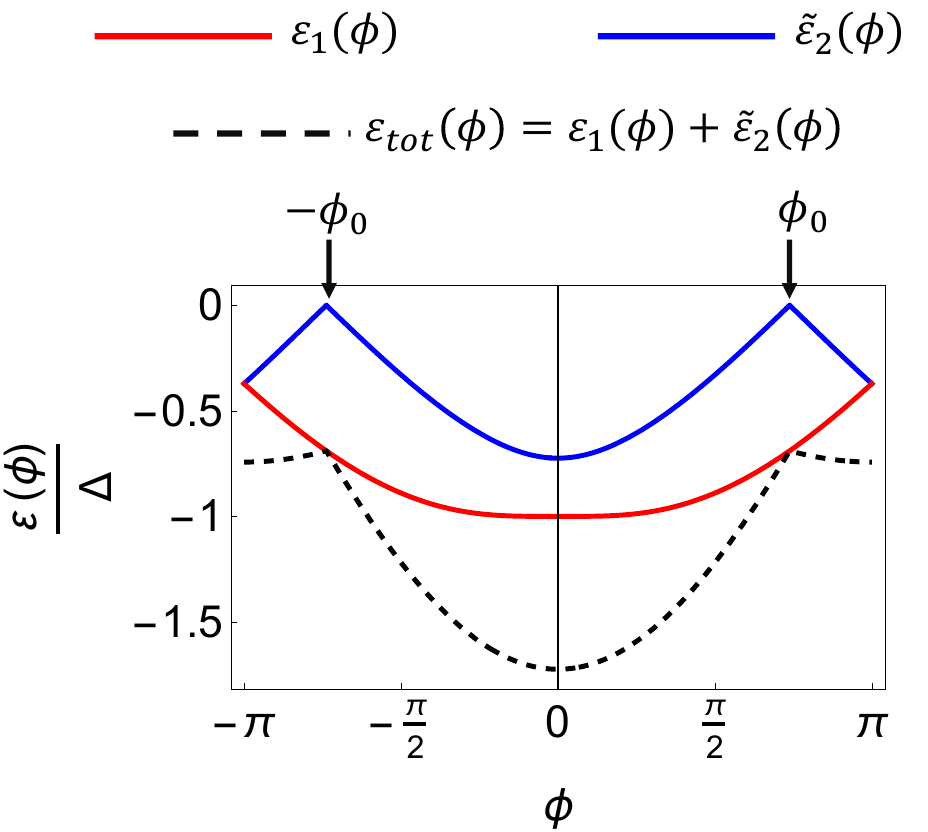}
		\caption{Current carrying Andreev levels and the total energy state, defined as the sum of these two levels, that effectively carries current through the SQDS JJ.}
		\label{Fig_14_A_2_Andreev_conduct_lev_scheme_ed_0_h_0_4_Delta_0_02}
	\end{figure}

	Since the current is expressed in terms of the negative energy part of the Andreev spectrum we can define the current-carrying ABS level as follows
	
	\begin{equation}
		\epsilon_{tot}(\phi)=\epsilon_{1}\left(\phi\right)+\tilde{\epsilon}_{2}\left(\phi\right)
		\label{Total_ABS_energy}
	\end{equation}
	where $\tilde{\epsilon}_{2}\left(\phi\right)$ reads
	\begin{equation}
		\begin{split}
		&\tilde{\epsilon}_{2}\left(\phi\right)=\epsilon_{3}\left(\phi\right) \left[\Theta\left(-\phi-\phi_{0}\right)+\Theta\left(\phi-\phi_{0}\right)\right]+ \\
		&\epsilon_{2}\left(\phi\right) \left[1-\Theta\left(-\phi-\phi_{0}\right)-\Theta\left(\phi-\phi_{0}\right)\right],
	\end{split}
	\label{Current_carrying_ABS}
	\end{equation}
	and it is reported in Fig.\ref{Fig_14_A_2_Andreev_conduct_lev_scheme_ed_0_h_0_4_Delta_0_02}.
	
	Expression for $\tilde{\epsilon}_{2}\left(\phi\right)$ can be simplified by noticing that the two crossing levels are particle-hole symmetric, thus, $\epsilon_{3}\left(\phi\right)=-\epsilon_{2}\left(\phi\right)$, leading to
	
	\begin{equation}
		\tilde{\epsilon}_{2}\left(\phi\right)=\epsilon_{2}\left(\phi\right) \left[1-2\Theta\left(-\phi-\phi_{0}\right)-2\Theta\left(\phi-\phi_{0}\right)\right].
		\label{level_def}
	\end{equation}
	When computing the Andreev bound states contribution to the current, we have to perform the energy derivative along $\phi$, according to the equation:
	
	\begin{equation}
		J(\phi)=\partial_{\phi}\epsilon_{tot}\left(\phi\right),
	\end{equation}
	that, using the expression in Eq. \ref{level_def}, yields
	
	\begin{equation}
		\begin{split}
			J(\phi)
			=&\partial_{\phi}\epsilon_{2}\left[1-2\Theta\left(-\phi-\phi_{0}\right)-2\Theta\left(\phi-\phi_{0}\right)\right]+ \\
			&+\partial_{\phi}\epsilon_{1}+2\epsilon_{2} \left[\delta\left(\phi+\phi_{0}\right)-\delta\left(\phi-\phi_{0}\right)\right].\label{J_dphi_energy}
		\end{split}
	\end{equation}
	The first term in the right-hand side of Eq.\ref{J_dphi_energy} provides the current with jumps at $\phi=\pm \phi_{0}$, while second term vanishes, since the $\delta$ functions are identically $0$ except for the crossing points $\phi=\pm \phi_{0}$ where $\epsilon_{2}\left(\phi=\pm\phi_{0}\right)=0$. Therefore, the final expression for the CPR reads
	
	\begin{equation}
		J(\phi)=\partial_{\phi}\epsilon_{1}+\partial_{\phi}\epsilon_{2}\left[1-2\Theta\left(-\phi-\phi_{0}\right)-2\Theta\left(\phi-\phi_{0}\right)\right].
		\label{J_jumps}
	\end{equation}
	Starting from the CPR formula in Eq.\ref{J_jumps}, we can easily demonstrate the presence of $\delta$-like divergences in the  current derivatives along system parameters $\{\lambda_{i}\}$ (e.g. $h$) by direct calculation. 
	Performing the derivative along the system parameter $\lambda_{i}$, we get
	
\begin{equation}
	\begin{split}
	&\partial_{\lambda_{i}}J\left(\phi\right)=\partial_{\lambda_{i}}\partial_{\phi}\epsilon_{2}\left[1-2\Theta\left(-\phi-\phi_{0}\right)-2\Theta\left(\phi-\phi_{0}\right)\right]+\\
	&+\partial_{\lambda_{i}}\partial_{\phi}\epsilon_{1}+2\partial_{\phi}\epsilon_{2}\partial_{\lambda_{i}}\phi_{0} \left[\delta\left(\phi+\phi_{0}\right)-\delta\left(\phi-\phi_{0}\right)\right],
	\end{split}
	\label{dJ_ed_h_general}
\end{equation}
	where we use that $\phi_{0}$ is a function of $\{\lambda_{i}\}$ and consequently $\partial_{\lambda_{i}}\Theta\left(\pm\phi-\phi_{0}\right)$ yields
	
	\begin{equation}
		\partial_{\lambda_{i}}\Theta\left(\phi-\phi_{0}\right)=\partial_{\lambda_{i}}\phi_{0}\left(\{\lambda_{i}\}\right)\delta\left(\phi-\phi_{0}\right).
	\end{equation}

	In addition, the second term in the right-hand side of Eq.\ref{dJ_ed_h_general} describing the $\delta$ divergences does not vanish unless $\partial_{\lambda_{i}}\phi_{0}$ does, i.e. unless $\phi_{0}$ does not depend on the parameter $\lambda_{i}$.
	This demonstrates that, at $T=0$, the current derivatives along the dot energy and Zeeman field exhibits $\delta$-like divergences at the crossing points, $\phi=\pm\phi_{0}$, between two particle-hole symmetric Andreev levels.

\end{document}